\documentclass[11pt,a4paper,oneside]{amsart}

\usepackage{hyperref}

\usepackage[foot]{amsaddr}
\usepackage{amsmath}
\usepackage{amssymb}
\usepackage{amsthm}
\usepackage{cleveref}
\usepackage{comment}
\usepackage[usenames]{color}
\usepackage{enumitem}
\usepackage{geometry}
\usepackage{mathrsfs}
\usepackage{mathtools}
\usepackage[active]{srcltx}
\usepackage[normalem]{ulem}
\usepackage{aligned-overset}

\usepackage[
    type={CC},
    modifier={by},
    version={4.0},
    imagewidth={3.6em},
    imagemodifier={-80{x}15},
]{doclicense}

\theoremstyle{plain}
\newtheorem{lemma}{Lemma}[section]
\newtheorem{theorem}[lemma]{Theorem}

\theoremstyle{definition}

\newtheorem{definition}[lemma]{Definition}
\newtheorem{remark}[lemma]{Remark}
\newtheorem{example}[lemma]{Example}

\numberwithin{equation}{section}

\let\temp\phi%
\let\phi\varphi%
\let\temp\epsilon%
\let\epsilon\varepsilon%
\let\varepsilon\temp%

\DeclareMathOperator{\Dom}{Dom}
\DeclareMathOperator{\Ent}{Ent}

\DeclareMathOperator{\Length}{Length}

\DeclareMathOperator{\OptCaus}{OptCaus}
\DeclareMathOperator{\OptGeo}{OptGeo}
\DeclareMathOperator{\Caus}{Caus}

\DeclareMathOperator{\Ric}{Ric}
\DeclareMathOperator{\Scal}{Scal}

\DeclareMathOperator{\ee}{e}

\DeclareMathOperator{\vol}{Vol}

\newcommand{\de}{{\mathrm{d}}}

\newcommand{\CD}{\mathsf{CD}}

\newcommand{\RCD}{\mathsf{RCD}}

\newcommand{\Prob}{\mathcal{P}}

\newcommand{\mm}{\mathfrak{m}}

\newcommand{\sfd}{\mathsf{d}}

\newcommand{\staticWithoutH}{\mathfrak{s}}
\newcommand{\staticWithH}{\staticWithoutH(H)}

\newcommand{\initialWithoutH}{\mathfrak{a}}
\newcommand{\initialWithH}{\initialWithoutH(H)}

\newcommand{\finalWithoutH}{\mathfrak{b}}
\newcommand{\finalWithH}{\finalWithoutH(H)}

\setlist[enumerate]{leftmargin=1cm}
\setlist[itemize]{leftmargin=1cm}

\newcommand{\noending}[1]{{#1}^\circ}

\title[The Null Energy Condition through the Lens of Optimal Transport]{The Null Energy Condition through the Lens of Optimal Transport: Smooth and Non-Smooth Spacetimes}

\author{Fabio Cavalletti}
\email{fabio.cavalletti@unimi.it}
\author{Davide Manini}
\email{davide.manini@univie.ac.at}
\author{Andrea Mondino}
\email{andrea.mondino@maths.ox.ac.uk}

\begin{document}

\begin{abstract}
We survey recent developments on the null energy condition (NEC) from the perspective of optimal transport, with particular emphasis on non-smooth Lorentzian geometry. After reviewing the role of classical energy conditions in general relativity, we discuss the connection between Ricci curvature and entropy convexity in optimal transport.

The main focus is the optimal-transport characterization of the NEC developed by the authors. In the smooth setting, the NEC is shown to be equivalent to the displacement concavity of the Shannon entropy power along suitable null-geodesic transport plans supported on null hypersurfaces and measured with respect to rigged measures. We review the geometric ingredients underlying this characterization, including null hypersurfaces, rigged measures, and optimal transport along null generators, together with applications such as a weighted Hawking's area theorem.

We then turn to the synthetic theory introduced in a second paper of the authors. Its starting point is a notion of synthetic null hypersurface based on achronal boundaries, gauge functions, and reference measures in topological causal spaces. Within this framework, the displacement-concavity characterization of the NEC is promoted to a synthetic curvature condition, denoted by \(\mathsf{NC}^{e}(N)\), which remains meaningful in the absence of any differentiable structure. We discuss its compatibility with the smooth theory, invariance under natural changes of gauge and reference measure, and stability under convergence. Finally, we present a synthetic version of Penrose's singularity theorem for continuous Lorentzian metrics. 

The results provide evidence that the geometric and causal consequences of the NEC persist far beyond the classical smooth setting and illustrate the potential of optimal transport as a unifying framework for curvature, causality, and singularity formation in Lorentzian geometry.

\end{abstract}


\maketitle
\vspace{-1.3 cm}
\tableofcontents

\section{Energy Conditions in General Relativity}

In general relativity, gravitation is described geometrically: the
gravitational field is encoded in the curvature of spacetime, while
matter and energy act as its source. More precisely, spacetime is
modelled by a Lorentzian manifold \((M,g)\), where \(M\) is a smooth,
connected manifold of dimension \(n\) (with physical dimension
\(n=4\)) and \(g\) is a Lorentzian metric, that is, a smooth,
symmetric, non-degenerate \((0,2)\)-tensor of signature \((- + \cdots +)\); see, e.g.,~\cite{HawEll,O'Neill}

The dynamics of the gravitational field are governed by Einstein's field equations
\begin{equation}\label{eq:Einstein}
G_{\mu\nu}+\Lambda g_{\mu\nu}=T_{\mu\nu},
\end{equation}
where
\[
G_{\mu\nu}:=\Ric_{\mu\nu}-\frac12\,\Scal\, g_{\mu\nu}
\]
is the Einstein tensor, \(\Ric\) and \(\Scal\) denote the Ricci tensor and scalar curvature of \(g\), respectively, \(\Lambda\in\mathbb{R}\) is the cosmological constant, and \(T_{\mu\nu}\) is the stress-energy tensor describing the local distribution of matter and energy. Using the contracted Bianchi identities, one obtains the conservation law \(\nabla^\mu T_{\mu\nu}=0\), expressing local conservation of energy and momentum.

A celebrated summary of Einstein's theory, often attributed to Wheeler~\cite{Gravitation}, is that ``matter tells spacetime how to curve, and curved spacetime tells matter how to move''. While the Einstein equations prescribe how geometry is related to matter, they impose surprisingly few intrinsic restrictions on the stress-energy tensor \(T_{\mu\nu}\). To exclude unphysical matter models and capture basic expectations about the positivity and causal propagation of energy, one therefore introduces additional assumptions known as \emph{energy conditions}.

Energy conditions are pointwise inequalities imposed on \(T_{\mu\nu}\) and are motivated by the requirement that observers should measure non-negative energy densities and that energy fluxes should not propagate faster than light. The most commonly studied examples are the null, weak, dominant, and strong energy conditions (NEC, WEC, DEC, and SEC). Although differing in their precise formulations, these conditions constrain the energy density and principal pressures measured by timelike or null observers~\cite{HawEll, Wald, Carroll}.

Historically, energy conditions have played a fundamental role in mathematical relativity. They constitute key hypotheses in the Hawking{--}Penrose singularity theorems~\cite{Penrose65, Haw:67}, Hawking's black-hole area theorem~\cite{Haw71}, topological censorship results~\cite{Penrose-DiffTopGR}. However, developments in quantum field theory and modern cosmology have revealed that several classical energy conditions can be violated, even in physically relevant situations such as Casimir systems, Hawking radiation, and models of accelerated cosmic expansion. As a consequence, the validity, interpretation, and possible replacements of classical energy conditions remain active topics of research in both mathematical and semiclassical gravity; for a recent survey, see~\cite{Curiel2017}.

We now recall the classical energy conditions. To state them precisely, we first fix some standard terminology from Lorentzian geometry. 

Let \((M,g)\) be a Lorentzian manifold and let \(p\in M\). A vector \(v\in T_pM\) is called
\[
\begin{cases}
\text{\emph{timelike}},\\
\text{\emph{null}} \; (\text{or \emph{lightlike}}),\\
\text{\emph{causal}},\\
\text{\emph{spacelike}},
\end{cases}
\qquad\text{according as}\qquad
g_p(v,v)
\begin{cases}
<0,\\
=0 \ \text{and}\ v\neq 0,\\
\le 0 \ \text{and}\ v\neq 0,\\
>0 \ \text{or}\ v=0.
\end{cases}
\]
respectively. The set of causal vectors in \(T_pM\) forms the \emph{causal cone}.
A Lorentzian manifold \((M,g)\) is said to be \emph{time-orientable} if there exists a continuous timelike vector field on \(M\). A choice of such a time orientation determines, at each point \(p\in M\), a decomposition of the causal cone into future and past components. A time-orientable Lorentzian manifold together with a fixed time orientation is called \emph{time-oriented} (also called \emph{a spacetime}).
Unless otherwise stated, all causal vectors appearing below are assumed to be future-directed.

Some classical energy conditions are:

\begin{itemize}
\item[\textbf{(WEC)}] \emph{Weak energy condition}: for every timelike vector \(v\in TM\),
\[
T(v,v)\ge 0.
\]
 Physically, this condition expresses the expectation that the effective energy density measured by massive observers should be non-negative.

\item[\textbf{(NEC)}] \emph{Null energy condition}: for every null vector \(v\in TM\),
\[
T(v,v)\ge 0.
\]
 Physically, this condition expresses the expectation that the effective energy density measured by photons moving at the speed of light should be non-negative.

\item[\textbf{(DEC)}] \emph{Dominant energy condition}:  for every
  future-directed timelike (or causal) vectors \(v,w\in TM\),
  \[
    T(v,w)\geq 0
    ,
  \]
  or, equivalently
  the vector field
\[
T(v):=-T^\mu{}_\nu v^\nu
\]
is causal and future-directed. In particular the DEC implies  the WEC.

\item[\textbf{(SEC)}] \emph{Strong energy condition}: for every timelike vector \(v\in TM\),
\[
\Ric(v,v)\ge 0.
\]
Although this interpretation should not be taken too literally, the SEC is often regarded in the physics literature as encoding the attractive nature of gravity for massive observers. The reason is that the condition $\Ric(v,v)\geq 0$ implies, via the Raychaudhuri equation, a focusing effect on timelike geodesic congruences, thereby tending to draw nearby freely falling observers together.

When a cosmological constant \(\Lambda\) is present, one often replaces this by the requirement
\[
\Ric(v,v)\ge -\frac{2\Lambda}{n-2}\, g(v,v),
\]
for all $v\in TM$ timelike,
which is the condition naturally arising from Einstein's equations.
\end{itemize}

\begin{figure}[ht]\label{fig1}
\caption{Summary of implications on energy conditions}
	\centering
  \includegraphics[scale=0.5]{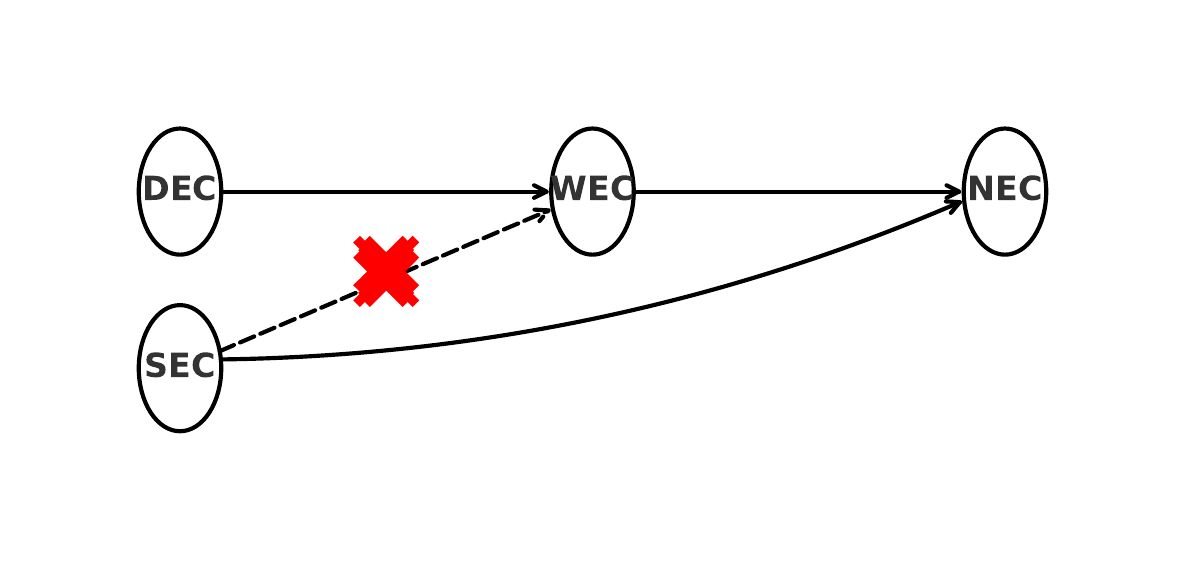}
\end{figure}

The logical implications among these conditions are summarized in Figure~\ref{fig1}.

\begin{example}[Energy conditions for a perfect fluid]
As a basic example, let us examine the classical energy conditions for a perfect fluid in four-dimensional spacetime. The stress-energy tensor of a perfect fluid is given by
\[
T_{\mu\nu}=(\rho+p)U_\mu U_\nu + p\, g_{\mu\nu},
\]
where \(\rho\) denotes the energy density measured in the fluid rest frame, \(p\) is the isotropic pressure, and \(U\) is the future-directed unit timelike four-velocity field, normalized by \(g(U,U)=-1\).

A straightforward computation shows that the classical energy conditions reduce to the following inequalities:

\begin{itemize}
\item[\textbf{(WEC)}:] \(\rho\ge 0\) and \(\rho+p\ge 0\);
\item[\textbf{(NEC)}:] \(\rho+p\ge 0\);
\item[\textbf{(DEC)}:] \(\rho\ge |p|\);
\item[\textbf{(SEC)}:] \(\rho+p\ge 0\) and \(\rho+3p\ge 0\).
\end{itemize}

Thus, for a perfect fluid, the energy conditions translate into simple constraints relating the energy density and pressure.
\end{example}

\begin{remark}
\begin{itemize}
\item Most classical matter models arising in physics satisfy the dominant energy condition. In particular, electromagnetic fields and many scalar-field models obey the DEC (and hence also the WEC and NEC).

\item The strong energy condition plays a central role in the Hawking{--}Penrose singularity theorems, as it guarantees the focusing of timelike geodesic congruences through the Raychaudhuri equation. Unlike the NEC, WEC, and DEC, however, the SEC is violated by several physically relevant matter models, including massive scalar fields.

\item A positive cosmological constant provides a particularly important example of SEC violation. Indeed, interpreting the cosmological constant as an effective perfect fluid with equation of state
\[
p=-\rho,
\]
one finds
\[
\rho+p=0,
\qquad
\rho+3p=-2\rho<0,
\]
for \(\rho>0\). Thus the NEC is saturated, while the SEC fails. This violation is closely related to the accelerated expansion of the universe.

\item Quantum field theories generally violate all of the classical pointwise energy conditions. Nevertheless, suitable averaged or integrated replacements often remain valid. Important examples include Ford's quantum inequalities~\cite{Ford1978}, their rigorous formulation in curved spacetimes~\cite{FewsterSmith2008}, and the quantum singularity theorems of Fewster and Galloway~\cite{FewsterGalloway2011}; for a recent survey see~\cite{KontouFewster2020}.
\end{itemize}
\end{remark}

Since the focus of this survey is the null energy condition, we begin by briefly recalling its geometric meaning and its role in some of the most fundamental results of general relativity.

\subsection*{The null energy condition}

Through Einstein's equations~\eqref{eq:Einstein}, the \emph{null energy condition} (NEC) can be expressed geometrically as
\begin{equation}\label{eq:NECIntro}
\Ric(v,v)\ge 0,
\qquad
\text{for every null (i.e.\ lightlike) vector } v.
\end{equation}

Among the classical energy conditions, the NEC is arguably the weakest and most robust. It is satisfied by essentially all standard classical matter models and occupies a distinguished position in Lorentzian geometry because it governs the focusing of null geodesics through the Raychaudhuri equation; see, for instance,~\cite[Sec.~9.2]{Wald}. As a consequence, the NEC plays a central role in a wide range of geometric, causal, and physical results.

Two particularly notable applications are the following.

\begin{enumerate}
\item \emph{Hawking's area theorem.}
The NEC is a key ingredient in the proof of Hawking's area theorem~\cite{Haw71}, which states that the area of a black-hole event horizon cannot decrease in classical general relativity. This result provides a striking geometric analogue of the second law of thermodynamics and played a central role in the development of black-hole thermodynamics.

\item \emph{Penrose's singularity theorem.}
The NEC is one of the fundamental assumptions in Penrose's singularity theorem~\cite{Penrose65}. Under suitable global and causal hypotheses, together with the existence of a trapped surface, the theorem implies causal geodesic incompleteness, typically interpreted as the formation of a spacetime singularity. This result marked the birth of modern singularity theory and remains one of the cornerstones of mathematical general relativity.
\end{enumerate}

Because of its intimate connection with the focusing of null geodesics, trapped surfaces, and singularity formation, the NEC occupies a distinguished role among the classical energy conditions. Nevertheless, the synthetic theory of the NEC is most naturally understood against the backdrop of earlier developments concerning timelike Ricci curvature and the {SEC}. Accordingly, we first briefly review the synthetic timelike theory before turning to a more detailed discussion of recent synthetic formulations of the NEC in non-smooth spacetime settings.

\section{Non-smooth spacetimes, an overview}

\subsection{Why going beyond smooth spacetimes}\label{sec:2.1}

Already in the 1950s, Lichnerowicz emphasized that physically realistic models in general relativity naturally lead to spacetime metrics of limited regularity~\cite{Lichn}. Indeed, many matter models give rise to stress-energy tensors that are not smooth, and the Einstein equations then imply that the Lorentzian metric itself may fail to be \(C^2\), the minimal regularity class required for a classical definition of the curvature tensor. Such low-regularity phenomena occur in a wide variety of physically relevant situations, including matched spacetimes describing stellar interiors and exteriors~\cite{MaSe}, self-gravitating compressible fluids and shock waves~\cite{LeFlochMardare2007, BuLe}, impulsive gravitational waves~\cite{PenGW}, and more recently in models of cyclic cosmology~\cite{LLV}.

A major breakthrough in the mathematical treatment of such spacetimes was achieved by Geroch and Traschen~\cite{GeTr}, who showed that if
\[
g\in L^\infty_{\mathrm{loc}}\cap W^{1,2}_{\mathrm{loc}},
\qquad
g^{-1}\in L^\infty_{\mathrm{loc}},
\]
then the Riemann and Ricci curvature tensors can be interpreted as distributions, and Einstein's equations retain a well-defined distributional meaning. This framework has since become a cornerstone of the theory of low-regularity Lorentzian geometry.

A fundamental feature of the Geroch{--}Traschen approach is that the underlying spacetime manifold \(M\) remains a smooth differentiable manifold, while only the metric tensor is allowed to lose regularity. This naturally raises a more radical question: can one formulate a meaningful notion of spacetime geometry when the smooth manifold structure itself is absent or ill-defined? Such a question is motivated not only by the study of singularities, where classical general relativity is expected to break down, but also by several approaches to quantum gravity. In many candidate theories, smooth spacetime is no longer regarded as fundamental; rather, it is expected to emerge only as a macroscopic approximation of a more primitive discrete, combinatorial, or quantum-geometric structure at Planckian scales.

\subsection{Synthetic approaches to non-smooth spacetimes}

A useful analogy may be drawn with the theory of fluids. At macroscopic scales, a fluid is well described by continuum models governed by partial differential equations, despite the fact that matter is fundamentally composed of discrete molecules. Classical solutions of the governing equations often develop singularities, such as shocks or concentration phenomena, beyond which smoothness breaks down. Rather than abandoning the continuum description altogether, one enlarges the class of admissible solutions by introducing weak functional frameworks, such as Sobolev spaces, distributions, and functions of bounded variation. This strategy has proved remarkably successful: singular phenomena can be accommodated and analyzed within a continuous theory, even though their ultimate physical resolution may require a more microscopic description.

A similar perspective can be adopted in general relativity. Classical spacetime models are based on smooth Lorentzian manifolds, yet both physical considerations and mathematical results suggest that singularities are unavoidable under rather general assumptions~\cite{HawkingPenrose1970}. One possible response is to seek a more fundamental theory of quantum gravity in which the smooth spacetime picture is replaced by a discrete or quantum structure. A complementary approach, which is the focus of the present discussion, is to enlarge the class of admissible spacetime geometries and develop a weak or synthetic theory capable of describing singular spacetimes while remaining within a geometric framework.

Several such theories have emerged.
Notable examples include the theory of timelike spaces by Busemann~\cite{Busemann}, closed ordered spaces by Nachbin~\cite{Nachbin},  causal spaces by Kronheimer and Penrose~\cite{CausalSpace}, and  
Lorentzian {(pre-)}length spaces  by Kunzinger and S\"amann~\cite{KS}. At present, several variants of the basic axiomatization of these spaces have been developed; see~\cite{McCann-NEC, Braun-McCann, Octet, Minguzzi-Suhr, ByMiSu:2025}. These frameworks dispense with the differentiable structure traditionally assumed in Lorentzian geometry and retain only the causal and metric features that are essential for spacetime theory.

A fundamental challenge immediately arises. In vacuum, Einstein's equations take the form
\[
\Ric \equiv 0,
\]
so that spacetime geometry is governed by Ricci curvature. Classically, however, Ricci curvature is defined through second derivatives of the metric tensor. If the underlying space is no longer a differentiable manifold, the very notion of differentiating the metric ceases to make sense. Consequently, one must find alternative ways of encoding curvature without relying on smooth differential calculus.

\subsection{Optimal transport and synthetic Ricci curvature lower bounds}

\subsubsection{Riemannian signature}
The connection between Ricci curvature and optimal transport originates in Riemannian geometry. A series of seminal works by McCann~\cite{McCann}, Otto{--}Villani~\cite{OttoVillani}, Cordero-Erausquin, McCann, and Schmuckenschl\"ager~\cite{CEMS}, and von Renesse{--}Sturm~\cite{vRS} established that lower Ricci curvature bounds can be characterized through displacement convexity properties of suitable entropy functionals along Wasserstein geodesics. This discovery culminated in the celebrated Lott{--}Sturm{--}Villani theory~\cite{lottvillani,sturm:I,sturm:II}, which introduced the curvature-dimension condition \(\CD(K,N)\), providing a synthetic notion of a metric measure space with Ricci curvature bounded below by \(K\in\mathbb R\) and dimension bounded above by \(N\in[1,\infty]\).

One of the most remarkable features of the \(\CD(K,N)\) theory is that curvature is encoded entirely through optimal transport and entropy convexity, without relying on any differentiable structure. As a consequence, meaningful notions of Ricci curvature extend far beyond the class of smooth Riemannian manifolds.

A major advance came with the work of Ambrosio, Gigli and Savar\'e~\cite{AGS14a,AGS14b}, who uncovered a deep connection between lower Ricci curvature bounds and gradient-flow structures. More precisely, they proved that in \(\CD(K,\infty)\) spaces the heat flow can be characterized as the \(W_2\)-Wasserstein gradient flow of the Boltzmann{--}Shannon entropy. Even more significantly, they showed that the associated gradient flow satisfies the Evolution Variational Inequality (EVI) formulation if and only if the heat flow is linear (see also the subsequent generalizations~\cite{AGMR, AES:16}). This observation led to the identification of the genuinely ``Riemannian'' synthetic structures within the larger class of \(\CD\) spaces, which also includes Finsler geometries.

These ideas gave rise to the theory of \(\RCD\) spaces, whose fundamental additional requirement is the infinitesimal Hilbertianity of the Sobolev space \(W^{1,2}\). The analytic and geometric consequences of this condition were systematically developed by Gigli~\cite{Gigli15}, leading to a rich first- and second-order differential calculus in the absence of smoothness.
A fundamental property of the $\RCD$ condition is the stability under pointed measured Gromov-Hausdorff convergence, proved by Gigli, Mondino and Savar\'e~\cite{GMS15}.

A further milestone was the realization that the optimal transport formulation of lower Ricci bounds is equivalent to the Eulerian viewpoint based on Bochner-type inequalities, also known  as the Bakry--\'Emery condition. This equivalence was first established in the infinite-dimensional setting by Ambrosio, Gigli and Savar\'e~\cite{AGS15}, and subsequently in finite dimensions by Erbar, Kuwada and Sturm~\cite{EKS} and, independently, by Ambrosio, Mondino and Savar\'e~\cite{AMS19}. Another cornerstone of the theory is the local-to-global property of the \(\RCD(K,N)\) condition, established by Cavalletti and Milman~\cite{CaMi:21}.

The resulting \(\RCD(K,N)\) theory has undergone spectacular development over the last decade and now provides one of the most powerful frameworks for geometric analysis in non-smooth spaces. We refer to the surveys~\cite{Vil:BS,Amb:ICM,Sturm:ECM,G:Survey} for comprehensive introductions and further references.

\subsubsection{Timelike Ricci lower bounds}
More recently, analogous ideas have been developed in Lorentzian geometry. In contrast to the Riemannian setting, the presence of a causal structure fundamentally alters the nature of the transport problem: admissible transport plans must respect causality, and the resulting geometry is inherently asymmetric in time. In the timelike setting, McCann~\cite{McCann} obtained an optimal transport characterization of the strong energy condition. Independently, and shortly thereafter, Mondino and Suhr~\cite{MoSu} showed that lower bounds on timelike Ricci curvature—and, more generally, the Einstein equations themselves—can be reformulated through displacement convexity properties of suitable entropy functionals along causal transport geodesics. These works established a conceptual bridge between Lorentzian curvature and optimal transport, providing the smooth foundation for the synthetic theory of timelike Ricci curvature bounds developed by Cavalletti and Mondino~\cite{CaMo:20} within the framework of Lorentzian {(pre-)}length spaces introduced by Kunzinger and S\"amann~\cite{KS}.

One of the main achievements of the theory developed in~\cite{CaMo:20} (see also the survey~\cite{CM:22}) is the extension of Hawking's singularity theorem to a fully synthetic setting. Remarkably, the resulting theorem requires neither a differentiable structure nor a smooth Lorentzian metric, thereby showing that singularity formation can be understood as a genuinely synthetic phenomenon. Closely related developments include inextendibility results for Lorentzian pre-length spaces satisfying synthetic lower bounds on timelike sectional curvature~\cite{GKS:19,AGKS:23,BHS:26}, as well as singularity theorems in the general setting of closed cone structures~\cite{Min}.  Together, these works provide strong evidence that several fundamental features of Lorentzian geometry persist far beyond the classical smooth framework.

More broadly, this line of research has led to synthetic counterparts of a number of classical comparison, rigidity, and singularity theorems from Lorentzian geometry; see, for instance,~\cite{Braun,Octet,Braun-McCann,Braun:exact,BR:GL}. In addition, the optimal transport approach has recently yielded new Lorentzian isoperimetric and geometric inequalities under synthetic timelike Ricci curvature lower bounds~\cite{CM24-IsopLor}, highlighting the richness of the emerging synthetic theory.

A notable feature of the synthetic framework is that it accommodates singular spacetimes lying beyond the scope of classical Lorentzian geometry. In particular, the class of Lorentzian pre-length spaces with synthetic timelike Ricci curvature bounded below includes physically significant examples such as Penrose's impulsive gravitational waves~\cite{MRS:IPPW} and spacetimes endowed with Lipschitz Lorentzian metrics whose timelike Ricci curvature is bounded below in the sense of distributions~\cite{BSC:Lipsch}. This demonstrates that synthetic timelike Ricci bounds are sufficiently robust to capture geometries with genuinely low regularity while retaining meaningful curvature information.

\subsubsection{Ricci lower bounds in null directions}
While timelike Ricci curvature bounds provide a synthetic counterpart of the strong energy condition, a natural and equally important question is whether curvature bounds in null directions admit a similar formulation. From the perspective of general relativity, this amounts to developing a synthetic theory of the null energy condition (NEC), one of the most fundamental assumptions in causal and singularity theory. Since this topic constitutes the main focus of the next sections, we conclude this overview by briefly summarizing the principal developments in the area.

A first synthetic characterization of the NEC for smooth spacetimes was obtained by McCann~\cite{McCann-NEC}. The key idea is to view the NEC as a limiting case of timelike Ricci curvature lower bounds and to exploit the corresponding optimal transport formulations developed in~\cite{McCann,MoSu}. This approach is closely related to the synthetic \(\mathsf{TCD}(K,N)\) theory introduced in~\cite{CaMo:20} for Lorentzian length spaces. A notable feature of McCann's characterization is that it naturally extends to non-smooth Lorentzian length spaces. However, as already observed in~\cite{McCann-NEC}, the resulting condition is not stable under suitable notions of convergence.

A different perspective was subsequently introduced by Ketterer~\cite{Ket24}. Rather than approaching the NEC through a limiting procedure from timelike directions, Ketterer directly investigated optimal transport along null hypersurfaces. More precisely, he established a characterization of the NEC in terms of displacement convexity properties of entropy functionals for singular probability measures concentrated on codimension-two spacelike submanifolds contained in null hypersurfaces.

The authors~\cite{CMM24a} later obtained an optimal transport characterization of the NEC based on \emph{diffuse} probability measures supported on null hypersurfaces. More precisely, they showed that the NEC is equivalent to suitable entropy convexity inequalities along null transport geodesics. Their work develops a systematic framework linking Ricci curvature in null directions, rigged measures on null hypersurfaces, and optimal transport of probability measures that are absolutely continuous with respect to such measures. A key advantage of the diffuse-measure approach over the singular-measure framework of~\cite{Ket24} is its compatibility with stability and convergence properties. This makes it particularly suitable for the development of a synthetic theory and ultimately yields the stable synthetic formulation of the null energy condition obtained in~\cite{CMM:25}. The main achievements of~\cite{CMM:25}, which form the basis of the subsequent sections, are fourfold:
\begin{itemize}
\item the introduction of well-posed and geometrically natural synthetic notions of null Ricci curvature lower bounds and of the null energy condition;

\item the proof that these synthetic notions are fully compatible with the classical smooth theory;

\item the establishment of stability under a suitable notion of convergence for synthetic null hypersurfaces, thereby overcoming a major limitation of previous approaches;

\item the extension of fundamental results of Lorentzian geometry, including Hawking's area theorem and Penrose's singularity theorem, to highly non-smooth settings.
\end{itemize}

\subsection{Why extending singularity theorems beyond the smooth setting?}

The singularity theorems of Penrose~\cite{Penrose65} and Hawking~\cite{Haw:67} are among the most celebrated achievements of twentieth-century mathematical physics. They show that spacetime singularities---understood in terms of causal geodesic incompleteness---are not artifacts of highly symmetric solutions, but rather arise generically under physically natural assumptions. In particular, Penrose's theorem predicts singularity formation in gravitational collapse, while Hawking's theorem applies to cosmological models undergoing expansion. Together with their subsequent generalizations, these results profoundly shaped modern understanding of black holes and the evolution of the universe.

The classical proofs, however, rely on a smooth Lorentzian metric, typically of class \(C^2\) or higher, so that curvature tensors and the Raychaudhuri equation are well defined. This raises a natural question: to what extent is smoothness essential to the phenomenon of singularity formation? From both a mathematical and a physical perspective, there are compelling reasons to seek singularity theorems under substantially weaker regularity assumptions.

Indeed, many physically relevant spacetime models possess limited regularity, see Section~\ref{sec:2.1}.  In such situations, the metric may fail to be \(C^2\), while still retaining clear geometric and physical significance. Consequently, if singularity theorems were valid only in the smooth category, their physical relevance would be considerably diminished.

This issue was already emphasized by Hawking and Ellis in their seminal monograph~\cite{HawEll}. As they observed, the robustness of the singularity theorems requires that their conclusions persist under realistic reductions in regularity. Otherwise, geodesic incompleteness could in principle be avoided by a mild loss of differentiability rather than by any genuinely geometric mechanism. From the perspective of Einstein's equations, this would be particularly unsatisfactory, since metrics of regularity below \(C^2\) naturally arise from matter fields of limited regularity and do not necessarily correspond to physically pathological situations.

Although these concerns have been present since the early days of singularity theory (see, e.g.,~\cite[Sec.~6.2]{Sen:98}), substantial progress has only been achieved in recent years. On the one hand, Lorentzian causality theory has been successfully extended to continuous metrics~\cite{Chr-Grant,SaC0} and, more generally, to synthetic settings such as Lorentzian {(pre-)}length spaces and closed cone structures~\cite{KS,Min}. On the other hand, new analytical techniques, notably regularization methods adapted to low-regularity Lorentzian geometry, have led to versions of Penrose's singularity theorem for \(C^1\) metrics~\cite{Graf} and of Hawking's singularity theorem for \(C^{0,1}\) metrics~\cite{CGHKS-25}.

An even more ambitious goal is to formulate singularity theorems in settings where no differentiable structure is available. In this direction,~\cite{CaMo:20} established a synthetic version of Hawking's singularity theorem for Lorentzian length spaces satisfying suitable synthetic timelike Ricci curvature lower bounds; see also Braun{--}McCann~\cite{Braun-McCann} for extensions to variable lower bounds. These results demonstrate that singularity formation can be understood as a genuinely synthetic phenomenon, independent of smooth differential geometry.

Extending Penrose's theorem to a similarly low-regularity or synthetic framework is considerably more subtle. Unlike Hawking's theorem, which is governed by timelike Ricci curvature bounds, Penrose's theorem is controlled by curvature in null directions and is therefore intrinsically linked to the null energy condition. As a consequence, the development of synthetic notions of null Ricci curvature and the corresponding singularity theory has become one of the central challenges in modern non-smooth Lorentzian geometry. Next sections describe recent progress in this direction, culminating in the extension of Penrose's singularity theorem to continuous spacetimes established in~\cite{CMM:25}.

\section{Characterization of the null energy condition via optimal transport in smooth spacetimes}
In this section we review some of the main results of~\cite{CMM24a}.

\subsection{Some basics of Lorentzian geometry}
Let \((M,g)\) be a time-oriented Lorentzian manifold of dimension \(n\ge 2\). Fix an auxiliary smooth Riemannian metric on \(M\); all notions of local Lipschitz regularity are understood with respect to this metric (and hence with respect to any such metric).

A locally Lipschitz curve \(\gamma:I\to M\) is said to be \emph{causal} if its tangent vector \(\dot\gamma\) is future-directed and causal at almost every point where it is defined. It is said to be \emph{timelike} (or \emph{chronological}) if, in addition,
\[
g(\dot\gamma,\dot\gamma)<0
\qquad\text{a.e.\ on }I.
\]
The spacetime \((M,g)\) is called \emph{causal} if it contains no closed causal curves.

A causal curve is said to be \emph{inextendible} if its domain cannot be enlarged while preserving causality. Given two points \(x,y\in M\), we write
\[
x\le y
\qquad\text{(respectively, }x\ll y\text{)}
\]
if there exists a causal (respectively, timelike) curve connecting \(x\) to \(y\). 
The causal (resp.\ chronological) future of a point $p\in M$ is denoted as
\begin{align*}
J^+(p)&:=\{q\in M: p\leq q\}\\
I^+(p)&:=\{q\in M: p\ll q\}.
\end{align*}
An important role will also be played by the causal relation
\[J:=\{(p,q)\in M\times M \colon p\leq q\}.\]

The \emph{Lorentzian length} of a causal curve \(\gamma:I\to M\) is defined by
\[
\Length(\gamma)
:=
\int_I \sqrt{-g(\dot\gamma_t,\dot\gamma_t)},dt,
\]
and the associated \emph{time-separation function} (or \emph{Lorentzian distance}) is
\begin{equation}\label{eq:deftau}
\ell(x,y)
:=
\begin{cases}
\sup\{\Length(\gamma):\gamma \text{ causal from }x\text{ to }y\},
& x\le y,\\
-\infty,
& \text{otherwise}.
\end{cases}
\end{equation}
Unlike the Riemannian distance, the function \(\ell\) is not symmetric in general. Nevertheless it satisfies the reversed triangle inequality on causal triples:
\begin{equation}\label{eq:tauRevTr}
\ell(x,y)+\ell(y,z)\leq \ell(x,z), \quad \text{for all }x\leq y\leq z.
\end{equation}
We remark that the Lorentz distance is often defined to be $0$ if $x$
and $y$ are causally unrelated; however, the convention in this paper
(and in many papers in synthetic Lorentzian geometry) is more convenient
for an optimal transport perspective.

Finally, a causal curve \(\gamma:I\to M\) is called a \emph{geodesic} if it satisfies
\[
\nabla_{\dot\gamma}\dot\gamma=0.
\]
Equivalently, geodesics are locally represented by the exponential map and arise as the critical points of the Lorentzian length functional.

\subsection{Null hypersurfaces}
A central role in the study of the null energy condition is played by \emph{null hypersurfaces}. Recall that a hypersurface $H\subset M$ is called \emph{null} if the restriction $g_H:=g|_{TH}$ is degenerate. Equivalently, for every $p\in H$, there exists a non-zero vector $L_p\in T_pH$ such that
\[
g(L_p,w)=0
\qquad\text{for all }w\in T_pH.
\]
The collection of such vectors forms a one-dimensional distribution, called the \emph{null direction} of $H$.

The geometry of null hypersurfaces differs substantially from that of spacelike  hypersurfaces. First, since the induced metric $g_H$ is degenerate, its determinant vanishes identically, and therefore $H$ carries no canonical volume form induced by the ambient Lorentzian metric. Second, the Lorentzian distance becomes trivial along $H$: locally, any two points $x,y\in H$ are either causally unrelated or satisfy
\[
\ell(x,y)=0.
\]
As a consequence, neither the restricted volume form nor the Lorentzian distance provides a suitable measure-theoretic structure on $H$.

A key idea in the study of null hypersurfaces is that the degeneracy of $g_H$ can be compensated by selecting a null geodesic generator of the hypersurface. More precisely, let $L$ be a nowhere-vanishing null vector field tangent to $H$ satisfying
\[
g(L,L)=0,
\qquad
\nabla_L L =0.
\]
Such a vector field determines a natural measure on $H$, called the \emph{rigged measure} associated with $L$ and denoted by $\vol_L$. Although $g_H$ itself admits no volume form, the measure $\vol_L$ is non-trivial and is absolutely continuous with respect to any auxiliary $(n-1)$-dimensional Riemannian volume measure on $H$.

Heuristically, the construction amounts to replacing the degenerate null direction generated by $L$ with a direction of unit length and then computing the corresponding volume form in the usual way. Rigged measures provide the natural reference measures for the optimal transport theory on null hypersurfaces discussed in the next sections. 
The construction above relies on the classical rigging technique for null hypersurfaces. We refer to~\cite[Sect.~2.2.1]{CMM24a} for the version used in this context, building on the framework introduced in~\cite{Rigging}; see also~\cite[Sect.~2.6]{MinguzziCMP15} for a related exposition.

\subsection{Optimal transport on null hypersurfaces}

Having introduced rigged measures on null hypersurfaces, we now discuss the optimal transport framework developed in~\cite{CMM24a}, see also~\cite{Ket24} for related constructions. At first sight, transporting mass along a null hypersurface appears problematic. Indeed, the Lorentzian distance is completely degenerate on a null hypersurface, so that the standard Lorentzian optimal transport machinery cannot be applied directly. Nevertheless, the geometry of null hypersurfaces contains enough structure to support a meaningful transport theory.

Let \(H\subset M\) be a null hypersurface endowed with a rigged measure \(\vol_L\). Given two probability measures \(\mu_0,\mu_1\in\Prob(H)\), we consider the set of transport plans
\[
\Pi(\mu_0,\mu_1)
:=
\left\{
\pi\in\Prob(H\times H):
(P_1)_\sharp\pi=\mu_0,
(P_2)_\sharp\pi=\mu_1
\right\},
\]
where \(P_i\) denotes the projection onto the \(i\)-th factor. Among these, we distinguish the causally admissible plans
\[
\Pi_{\le}(\mu_0,\mu_1)
:=
\left\{
\pi\in\Pi(\mu_0,\mu_1):
\pi(J)=1
\right\}.
\]

We also introduce the class
\begin{equation}\label{eq:DefPac}
\Prob_{ac}(H)
:=
\{\mu\in\Prob(H):\mu\ll \vol_L\},
\end{equation}
which is independent of the particular choice of rigged measure.

In Lorentzian optimal transport~\cite{EM17, Suhr, McCann, MoSu, CaMo:20, OnSC:Orlicz}, the natural cost function is given by the time-separation,
\[
c(x,y)= \ell(x,y)
\]
or more generally by \(c(x,y)=\frac{1}{p}\ell(x,y)^p\) for $p\le 1$, $p\neq 0$. On a null hypersurface, however, this cost function becomes completely degenerate. Indeed, as observed above, for \(x,y\in H\) one has either \(x\not\le y\), in which case \(c(x,y)=-\infty\), or \(x\le y\), in which case
$
\ell(x,y)=0.
$
Thus the cost assumes only the values \(0\) and \(-\infty\).

At first sight, such a degenerate cost seems to destroy any meaningful transport theory. A crucial observation is that the set of points causally related to a given point \(x\in H\) is locally contained in a null generator of \(H\). In other words, the null geometry induces a natural partition of \(H\) into null geodesics, and transport can only occur along these one-dimensional leaves. This observation forms the starting point of the theory.

To formalize this idea, let
\[
\ee_t:C([0,1];H)\rightarrow H,
\qquad
\ee_t(\gamma):=\gamma_t,
\]
denote the evaluation map. 

\begin{definition}[Null-connected pairs of probability measures]
Two probability measures \(\mu_0,\mu_1\in\Prob(H)\) are said to be \emph{null connected along \(H\)} if there exists a probability measure
\[
\nu\in\Prob(C([0,1];H))
\]
such that:
\begin{itemize}
\item \(\nu\)-almost every curve \(\gamma\) is causal;
\item the associated endpoint coupling 
$
\pi:=(e_0,e_1)_\sharp\nu
$
belongs to \(\Pi_{\le}(\mu_0,\mu_1)\);
\item \(\ell(x,y)=0\) for \(\pi\)-almost every \((x,y)\in H\times H\).
\end{itemize}
The collection of all such dynamical transport plans is denoted by
$\OptCaus^H(\mu_0,\mu_1)$.
\end{definition}

Since the Lorentzian cost is degenerate, elements of \(\OptCaus^H(\mu_0,\mu_1)\) should be viewed as null analogues of optimal dynamical couplings. Among them, a distinguished role is played by those concentrated on null geodesics.

\begin{definition}
Let \(\mu_0,\mu_1\in\Prob(H)\) be null connected along \(H\). A dynamical transport plan $\nu\in\OptCaus^H(\mu_0,\mu_1)$
is called \emph{null-geodesic} if \(\nu\)-almost every curve is a null geodesic. We denote the collection of such plans by
$
\OptGeo^H(\mu_0,\mu_1).
$
\end{definition}
For a proof of the existence of null-geodesic transport plans see~\cite{CMM24a}:

\begin{theorem}\label{thm:ExistOptGeo}
Let \(\mu_0,\mu_1\in\Prob_{ac}(H)\) be such that \(\mu_0\) is null connected to \(\mu_1\) along \(H\). Then
\[
\OptGeo^H(\mu_0,\mu_1)\neq\emptyset.
\]
Equivalently, there exists a dynamical transport plan connecting \(\mu_0\) and \(\mu_1\) which is concentrated on null geodesics.
\end{theorem}

This result provides the basic geometric object required for the synthetic study of curvature in null directions. Indeed, once a suitable notion of geodesic transport on a null hypersurface is available, one can investigate the behaviour of entropy functionals along such transports and thereby relate their convexity properties to the null energy condition. 

The degeneracy of the cost function on a null hypersurface implies that the set $\OptGeo^H(\mu_0,\mu_1)$ is typically formed by more than one element. This phenomenon reflects the fact that all causally related points on $H$ have vanishing Lorentzian distance. Nevertheless,~\cite{CMM24a} shows that uniqueness can be recovered by selecting transport plans satisfying suitable monotonicity conditions with respect to the causal relation $J$; see~\cite[Sect.~6.1]{CMM24a}.

\subsection{Null energy condition and convexity properties of the  entropy along null optimal transport}\label{SS:NEC-OT-smooth}
We now turn to the central ingredient in the optimal transport characterization of the null energy condition, namely the displacement convexity of entropy functionals along null-geodesic transport plans. The results discussed in this section originate from~\cite[Sect.~7]{CMM24a}. Here we present a streamlined account tailored to the subsequent development of the non-smooth theory.

\subsubsection{Entropy functionals}

To formulate the relevant entropy inequalities, let \((X,\mathfrak m)\) be a measure space. The \emph{Boltzmann{--}Shannon entropy} of a probability measure \(\mu\in\Prob(X)\) with respect to the reference measure \(\mathfrak m\) is defined by
\[
\Ent(\mu|\mathfrak m)
:=
\int_X \rho\log(\rho)\,\de \mathfrak m,
\]
provided \(\mu=\rho\,\mathfrak m\) and \((\rho\log\rho)_+\in L^1(\mathfrak m)\); otherwise one sets
\[
\Ent(\mu|\mathfrak m):=+\infty.
\]
Originally introduced by Boltzmann in statistical mechanics and later rediscovered by Shannon in information theory, this functional has become a central object linking geometry, probability, analysis, and optimal transport.

We denote by
\[
\Dom(\Ent(\cdot|\mathfrak m))
:=
\{\mu\in\Prob(X):\Ent(\mu|\mathfrak m)<+\infty\}
\]
the effective domain of the entropy.

Following~\cite{EKS}, it is convenient to consider the dimensional entropy-power functional. Given \(N>0\), define
\[
\mathcal U_N(\mu|\mathfrak m)
:=
\exp\!\left(
-\frac{\Ent(\mu|\mathfrak m)}{N}
\right),
\qquad
\mu\in\Dom(\Ent(\cdot|\mathfrak m)),
\]
and set \(\mathcal U_N(\mu|\mathfrak m):=0\) otherwise.

The functional \(\mathcal U_N\) is closely related to the classical Shannon entropy power from information theory (see, e.g.,~\cite{DCT-1991}); for instance, in Euclidean space \(\mathbb R^N\) the entropy power corresponds to \(\mathcal U_{N/2}\). Its relevance in geometry stems from the fact that convexity properties of \(\mathcal U_N\) along optimal transport geodesics provide a synthetic formulation of lower Ricci curvature bounds. As we shall see, an analogous phenomenon occurs in the Lorentzian setting when optimal transport is performed along null hypersurfaces.

\subsubsection{The null energy condition as displacement concavity}

We are now in a position to state the main result of~\cite{CMM24a}, which provides an optimal transport characterization of the null energy condition. 

Let $H\subset M$ be a smooth null hypersurface and let $\nu\in \mathrm{OptGeo}^H(\mu_0,\mu_1)$
be a null-geodesic dynamical transport plan connecting two probability measures $\mu_0,\mu_1\in \mathcal{P}_{ac}(H)$. Denote by
\[
\mu_t:=(\ee_t)_\sharp \nu,
\qquad t\in[0,1],
\]
the associated interpolation, and define
\[
u_{n-1}(t):=\mathcal{U}_{n-1}(\mu_t \mid \mathrm{Vol}_L).
\]

The appearance of the exponent $n-1$ is natural, since a null hypersurface is an $(n-1)$-dimensional object and the corresponding entropy power is computed relative to the rigged measure $\mathrm{Vol}_L$.

The following theorem shows that the null energy condition is precisely equivalent to the displacement concavity of the entropy power along null-geodesic transports.

\begin{theorem}[\cite{CMM24a}]\label{thm:DisplConcNEC}
Let $(M^n,g)$ be a smooth spacetime. Then the following are equivalent:

\begin{enumerate}
\item The null energy condition holds, i.e.
\[
\operatorname{Ric}(v,v)\ge 0
\qquad
\text{for every null vector }v.
\]

\item For every smooth null hypersurface $H\subset M$, every pair of measures $\mu_0,\mu_1\in\mathcal{P}_{ac}(H)$ such that $\mu_0$ is null connected to $\mu_1$ along $H$, there exists a null-geodesic dynamical transport plan
\[
\nu\in \mathrm{OptGeo}^H(\mu_0,\mu_1)
\]
such that
\[
u_{n-1}(t)\ge (1-t)\,u_{n-1}(0)+t\,u_{n-1}(1),
\qquad
\forall\, t\in[0,1].
\]
\end{enumerate}
\end{theorem}

In other words, the null energy condition is equivalent to the concavity of the Shannon entropy power along suitable optimal transports contained in null hypersurfaces. This result should be viewed as the null counterpart of the displacement convexity characterizations of Ricci curvature lower bounds in Riemannian geometry.

The theorem admits a weighted generalization. Let
\[
\mathfrak{m}_L=e^\Phi\,\mathrm{Vol}_L
\]
be a smooth weighted measure on a null hypersurface, and let $N\ge n$. Consider the Bakry--\'Emery $N$-Ricci tensor
\[
\operatorname{Ric}^{g,\Phi,N}
:=
\operatorname{Ric}
-
\operatorname{Hess}\Phi
-
\frac{1}{N-n}\nabla\Phi\otimes\nabla\Phi.
\]
Then $\operatorname{Ric}^{g,\Phi,N}\ge 0$ in null directions if and only if the same displacement-concavity property holds with $\mathcal{U}_{N-1}$ in place of $\mathcal{U}_{n-1}$. Thus entropy power continues to encode weighted curvature bounds in exactly the same way as in the smooth Riemannian theory.

\subsubsection{The $\mathsf{NC}^e(N)$ condition}

One of the main advantages of the optimal transport formulation is that it remains meaningful under very weak regularity assumptions. In particular, the displacement-concavity inequalities above still make sense even when the weight function $\Phi$ is merely continuous.

Motivated by this observation,~\cite{CMM24a} introduced the following synthetic condition.

\begin{definition}
Let $(M^n,g)$ be a smooth spacetime, let $H\subset M$ be a smooth null hypersurface, and let
\[
\mathfrak{m}_L=e^\Phi\,\mathrm{Vol}_L
\]
with $\Phi\in C^0(H)$. Given $N\ge n$, we say that $(M,g,H,\Phi)$ satisfies the $\mathsf{NC}^e(N)$ condition if the entropy power $\mathcal{U}_{N-1}$ is displacement concave along null-geodesic transport plans in the sense of Theorem \ref{thm:DisplConcNEC}. Recall that the definition \eqref{eq:DefPac} of $\Prob_{ac}(H)$ is independent from the choice of $L$ and of the smooth weight $e^\Phi$.

We say that $(M,g,\Phi)$ satisfies the $\mathsf{NC}^e(N)$ condition if the above property holds for every smooth null hypersurface $H\subset M$.
\end{definition}

\begin{remark}
\begin{itemize}
\item Although the construction depends on the rigged measure $\mathrm{Vol}_L$, the resulting $\mathsf{NC}^e(N)$ condition is independent of the choice of null geodesic generator $L$. Thus it defines an intrinsic property of the spacetime (see~\cite[Remark 7.8]{CMM24a}).
\item The continuity assumption $\Phi\in C^0$ is substantially weaker than what would be required to interpret the Bakry--\'Emery tensor distributionally. Indeed, even defining
\[
\operatorname{Ric}^{g,\Phi,N}
=
\operatorname{Ric}
-
\operatorname{Hess}\Phi
-
\frac{1}{N-n}\nabla\Phi\otimes\nabla\Phi
\]
would typically require at least $\nabla\Phi\in L^2_{\mathrm{loc}}$. The optimal transport formulation therefore provides a genuine extension of weighted curvature bounds beyond the range accessible to distributional methods.

More importantly, the optimal transport formulation remains meaningful in situations where differential notions of curvature are unavailable. This observation motivates the synthetic theory developed in~\cite{CMM:25} and discussed in Section~\ref{Sec:NECSynt}, where the $\mathsf{NC}^e(N)$ condition is extended to non-smooth spacetimes and shown to be stable under suitable notions of convergence.
\end{itemize}
\end{remark}

\subsubsection{Geometric consequences}

Beyond providing a synthetic characterization of the null energy condition, the optimal transport approach developed in~\cite{CMM24a} yields several geometric and physical applications. Among the main consequences are:

\begin{itemize}

\item stability of the \(\mathsf{NC}^e(N)\) condition under natural approximation procedures, namely $C^1_{\rm loc}$-convergence of the Lorentzian metrics and $C^0_{\rm loc}$-convergence of the weights, see~\cite[Thm.\ 9.1]{CMM24a}. This robustness is one of the key features of the optimal transport formulation and foreshadows the stability results under measured Gromov{--}Hausdorff-type convergence obtained in~\cite{CMM:25} and described in the next section.

\item a weighted version of the classical light-cone theorem, together with a corresponding rigidity statement, extending earlier results of Choquet-Bruhat, Chru\'sciel and Mart\'{\i}n-Garc\'{\i}a~\cite{CBCMG-2009} and Grant~\cite{Grant}, see~\cite[Sect.\ 10.1]{CMM24a};

\item a weighted Hawking area theorem and an associated rigidity theorem.
\end{itemize}

We focus here on the weighted Hawking area theorem, which illustrates particularly well the geometric content of the \(\mathsf{NC}^e(N)\) condition.

Recall that a null hypersurface \(H\subset M\) is said to be \emph{future geodesically complete} if every future-directed null geodesic contained in \(H\) can be extended indefinitely towards the future. More precisely, whenever
\[
\gamma:[a,b]\to H
\]
is a future-directed null geodesic, there exists an extension
\[
\widetilde\gamma:[a,\infty)\to H
\]
which is again a future-directed null geodesic.

Let \(S\subset H\) be a spacelike cross-section of \(H\), i.e. \(H\) is spanned by null geodesics that pass through $S$ exactly once. The quantity
\[
\int_S e^\Phi\, \de\mathcal H^{n-2}
\]
may be interpreted as the weighted \((n-2)\)-dimensional area of \(S\) with respect to the density \(e^\Phi\). The following theorem asserts that such weighted areas are non-decreasing towards the future under the synthetic null energy condition.

\begin{theorem}[Weighted Hawking's area theorem {\cite[Thm.\ 10.3]{CMM24a}}]
Let \(H\subset M\) be a null hypersurface and let \(S_1,S_2\subset H\) be spacelike cross-sections satisfying
\[
S_1\subset J^-(S_2).
\]
Assume that \(g\), \(H\), \(S_1\), and \(S_2\) are of class \(C^2\). Let
\[
\mathfrak m=e^\Phi\,\mathrm{Vol}_L
\]
with \(\Phi:H\to\mathbb R\) continuous. If \(H\) is future geodesically complete and \((M,g,H,\Phi)\) satisfies the \(\mathsf{NC}^e(N)\) condition, then
\[
\int_{S_1} e^\Phi\,\de\mathcal H^{n-2}
\le
\int_{S_2} e^\Phi \, \de \mathcal H^{n-2}.
\]
\end{theorem}

When \(\Phi\equiv 0\), the theorem reduces to the classical monotonicity of the area of spacelike sections of a future-complete null hypersurface under the null energy condition discovered by Hawking~\cite{Haw71}. Thus the \(\mathsf{NC}^e(N)\) condition recovers one of the most important consequences of the NEC in black-hole geometry. Moreover, the theorem continues to hold for merely continuous weights, a level of regularity for which the Bakry--\'Emery Ricci tensor is not even defined in a distributional sense.

The proof illustrates a recurring theme in synthetic geometry: rather than studying the expansion of null geodesic congruences through differential equations such as the Raychaudhuri equation, one derives the desired monotonicity directly from entropy-concavity inequalities along null-geodesic transport plans. In this way, a classical geometric consequence of the null energy condition emerges from a purely optimal transport argument.

We refer to~\cite{CMM24a} for the corresponding rigidity statements, the proofs, and for the weighted light-cone theorem, which provide further evidence that the displacement-concavity approach captures the essential geometric content of the null energy condition.

\section{Synthetic null hypersurfaces and the null energy condition in the non-smooth setting}\label{Sec:NECSynt}

\subsection{The synthetic setting}

The developments described in the previous sections rely on the smooth structure of the underlying spacetime. A natural question is whether the optimal transport characterization of the null energy condition can be extended beyond the realm of smooth Lorentzian manifolds. More fundamentally, one may ask which aspects of general relativity remain meaningful when the differentiable structure itself is no longer available.

This perspective has deep roots in the foundations of the subject. In his 1969 lecture \emph{Gravitational Collapse: The Role of General Relativity}~\cite{Penrose:GRG02F}, Penrose argued that one should

\begin{quote}
``[{.{.}.}] examine, once more, the foundations of Einstein's theory, and to ask what parts of the theory are likely to be here to stay [{.{.}.}]. The parts of the theory I am referring to are, in fact, the geometrical interpretation of gravity, the curvature of space-time geometry, and general-relativistic causality.''
\end{quote}

Motivated by this philosophy, several authors have sought synthetic formulations of Lorentzian geometry in which the causal structure, rather than the differentiable structure, plays the fundamental role. A first step in this direction is to abstract the causal relations that are present in every spacetime.

\begin{definition}[Causal space]
A \emph{causal space} is a triple \((X,\le,\ll)\), where \(X\) is a set, \(\le\) is a preorder on \(X\), and \(\ll\) is a transitive relation satisfying
\(
\ll \;\subset\; \le.
\)
\end{definition}
The relations \(\le\) and \(\ll\) should be thought of as synthetic analogues of the causal and chronological relations of Lorentzian geometry. This notion originates in the causal spaces introduced by Kronheimer and Penrose~\cite{CausalSpace} and was later adopted by Kunzinger and S\"amann~\cite{KS} in their development of Lorentzian {(pre-)}length spaces.

In order to perform analysis and geometry, one must also endow the space with a suitable topology. The following class provided the natural framework for a the synthetic theory of the null energy condition developed in~\cite{CMM:25}.

\begin{definition}[Topological causal space]
A \emph{topological causal space} is a quadruple
\(
(X,\le,\ll,\mathfrak T)
\)
such that:

\begin{itemize}
\item \((X,\le,\ll)\) is a causal space;

\item \(\mathfrak T\) is a Polish and proper topology on (X);

\item the chronological relation \(\ll\) is open in \(X\times X\);

\item the causal relation \(\le\) is closed in \(X\times X\).
\end{itemize}
\end{definition}

These axioms capture some of the most basic topological and causal features of smooth spacetimes while making no reference to a differentiable structure, a metric tensor, or curvature. In particular, topological causal spaces belong to the broader class of closed ordered spaces studied by Nachbin~\cite{Nachbin}; see also the survey of Minguzzi~\cite{Minguzzi-Review2019} for a modern perspective.

The philosophy underlying the synthetic approach is that causality should be regarded as a primary geometric structure, from which increasingly refined notions of spacetime geometry can be reconstructed. The remainder of this chapter is devoted to explaining how the null energy condition can be formulated within this framework and how the resulting synthetic condition retains the stability and geometric consequences discussed in the smooth setting.

\subsection{Synthetic null hypersurfaces}

The smooth theory described in the previous sections relies heavily on the geometry of null hypersurfaces. In particular, the optimal transport characterization of the null energy condition is formulated in terms of entropy convexity along null-geodesic transport plans supported on a null hypersurface and measured relative to a rigged measure.

To extend this picture beyond the smooth category, one must first identify synthetic counterparts of the basic ingredients entering the construction. This raises a number of fundamental questions. How should one define a null hypersurface without a Lorentzian metric? What should replace the rigged measure? How can one distinguish the preferred affine parametrizations of null geodesics needed to formulate displacement convexity?

The purpose of this section is to explain how these issues are addressed in~\cite{CMM:25}.

\subsubsection{Achronal boundaries}

We begin with some standard terminology. Let \((X,\le,\ll,\mathfrak T)\) be a topological causal space. A subset \(S\subset X\) is said to be \emph{achronal} if
\[
x\not\ll y
\qquad
\forall,x,y\in S.
\]
Equivalently, no two points of \(S\) are in chronological relation.
\\Given a subset \(S\subset X\), its \emph{chronological future} is defined by
\[
I^+(S)
:=
\{y\in X:\exists\,x\in S \text{ such that } x\ll y\}.
\]
In a smooth spacetime, a null hypersurface can be characterized by the degeneracy of the induced metric. Since such a description is unavailable in the absence of a differentiable structure, a different viewpoint is required.

The key observation is that null hypersurfaces enjoy a strong causal property:  they are achronal (at least locally), yet the causal relation does not trivialize on them. This suggests replacing smooth null hypersurfaces by suitable achronal subsets. An especially important class is provided by achronal boundaries.

\begin{definition}
Let \((X,\le,\ll,\mathfrak T)\) be a topological causal space. A subset \(H\subset X\) is called an \emph{achronal boundary} if there exists a set \(S\subset X\) such that
\[
H=\partial I^+(S).
\]
\end{definition}

One can easily prove that achronal boundaries are indeed achronal
(see~\cite[Lem.~3.2]{CMM:25}).

The relevance of this notion stems from the smooth theory. Indeed, in smooth spacetimes, achronal boundaries are classical objects in causality theory (see for instance~\cite{Penrose-DiffTopGR, HawEll}). Under suitable assumptions they are ruled by null geodesics, and every maximal achronal set arises as an achronal boundary (see for instance~\cite[Thm.\ 2.97]{Minguzzi-Review2019}). Consequently, achronal boundaries provide a natural synthetic replacement for null hypersurfaces.

\subsubsection{Endpoints and static points of an achronal set} Although achronality is a fundamental property of null hypersurfaces, it is not sufficient to characterize them synthetically. Indeed, spacelike hypersurfaces are also achronal, but they do not possess the causal generators that are the hallmark of null geometry. To capture this additional structure, one must require that the causal relation remains non-trivial on the set. 
The next sections, culminating in
Definition~\ref{D:synthetic-null-hypersurface}, make this idea precise.

\begin{definition}\label{D:maxmin}
Let \(H\) be a closed achronal subset of a topological causal space.

The set of \emph{final points} of \(H\) is defined by
\[
\finalWithH
:=
\left\{
m\in H :
\nexists\,x\in H
\text{ such that }
m\le x
\text{ and }
x\neq m
\right\},
\]
that is, the points of \(H\) having no strictly larger causal successor in \(H\).

Similarly, the set of \emph{initial points} of \(H\) is defined by
\[
\initialWithH
:=
\left\{
m\in H :
\nexists\,x\in H
\text{ such that }
x\le m
\text{ and }
x\neq m
\right\},
\]
namely the points of \(H\) having no strictly smaller causal predecessor in \(H\).

The subset of \(H\) consisting of points that are neither initial nor final is denoted by
\begin{equation}\label{eq:defHnoend}
\noending{H}
:=
H\setminus
\bigl(
\initialWithH
\cup
\finalWithH
\bigr),
\end{equation}
and is called the \emph{achronal set without endpoints}.

Finally, the set of \emph{static points} of \(H\) is defined by
\[
\staticWithH
:=
\initialWithH
\cap
\finalWithH.
\]
Equivalently, a static point is a point of \(H\) that admits neither a non-trivial causal predecessor nor a non-trivial causal successor within \(H\).
\end{definition}

\subsubsection{Reference measures}

A second ingredient of the smooth theory is the rigged measure \(\mathrm{Vol}_L\), which serves as the reference measure in the definition of entropy.

Recall that, although the rigged measure depends on the choice of a null geodesic vector field \(L\), the resulting \(\mathsf{NC}^e(N)\) condition is independent of this choice. This suggests that the measure itself should be regarded as auxiliary data rather than as part of the intrinsic geometry.

In the synthetic setting, one therefore starts with a non-negative Radon measure
\[
\mathfrak m\in\mathcal M^+(H)
\]
defined on the achronal boundary \(H\). The measure \(\mathfrak m\) plays the role of the rigged measure and provides the reference measure appearing in the entropy functionals.

As in the smooth setting, the resulting synthetic curvature condition will ultimately be shown to be invariant under a suitable equivalence relation on reference measures. Roughly speaking, this equivalence corresponds to multiplying \(\mathfrak m\) by functions that remain constant along the null generators of \(H\), mirroring the change of rigging vector field in the smooth theory.

\subsubsection{Gauge functions and distinguished parametrizations}

A further ingredient is required before one can formulate a synthetic version of the \(\mathsf{NC}^e(N)\) condition. Recall that the latter is expressed through concavity properties of entropy along interpolating measures
$
(\mu_t)_{t\in[0,1]}
$
generated by null-geodesic transport plans. The parameter \(t\in [0,1]\) plays a crucial role, since concavity is a statement about the evolution of the entropy along the interpolation. Consequently, one needs a distinguished way of parametrizing the null geodesics supporting the transport.

In Riemannian geometry, every minimizing geodesic \(\gamma:[0,1]\to X\) admits a distinguished parametrization by constant speed, characterized purely in metric terms by the identity
\[
\sfd(\gamma_s,\gamma_t)
=
|t-s|\,\sfd(\gamma_0,\gamma_1),
\qquad
\forall\, s,t\in[0,1].
\]
This relation simultaneously encodes both the minimizing property of the curve and the fact that it is parametrized proportionally to arc length. Equivalently,
\[
|\dot\gamma|
\equiv
\Length(\gamma)
=
\sfd(\gamma_0,\gamma_1).
\]
An analogous phenomenon occurs in Lorentzian geometry: timelike geodesics admit a canonical parametrization by proper time, which can be characterized in terms of the time-separation function \(\ell\). Null geodesics are fundamentally different. For smooth spacetimes $(M,g)$, if $\gamma:[0,1]\to M$ is a null geodesic, any reparametrization still satisfies 
\[
g(\dot \gamma, \dot \gamma) \equiv \textrm{Length}(\gamma) = \ell (\gamma_{0},\gamma_{1}) = 0. 
\]
It follows that the class of maximizing null curves is invariant under arbitrary reparametrizations. As a consequence, the Lorentzian length functional does not determine a distinguished parametrization of null curves. This stands in sharp contrast with the Riemannian and timelike Lorentzian settings, where the length functional naturally selects constant-speed and proper-time parametrizations, respectively.

This becomes problematic because the \(\mathsf{NC}^e(N)\) condition is formulated through convexity properties of entropy along geodesics, and convexity is not preserved under arbitrary reparametrizations. In the smooth setting, one resolves this ambiguity by requiring null geodesics to satisfy the geodesic equation
\[
\nabla_{\dot\gamma}\dot\gamma=0,
\]
which fixes the parametrization up to affine transformations. Since convexity is invariant under affine reparametrizations, this is exactly the amount of structure needed.

The synthetic analogue of this affine parametrization is encoded by the notion of a gauge function. 
Intuitively, a gauge function plays the role of an affine parameter along null generators. In the smooth setting, one may think of \(G\) as a function satisfying
\[
G(\gamma_t)=t
\]
along suitably normalized null geodesics. 


\begin{definition}[Gauge function]
Let \(H\subset X\) be a closed achronal set, and let \(\noending{H}\) denote the corresponding achronal set without endpoints (see~\eqref{eq:defHnoend}).

A Borel function
\[
G:\noending{H}\to\mathbb R
\]
is called a \emph{gauge} for \(H\) if the following properties hold:

\begin{enumerate}
\item For every injective causal curve \(\gamma\) contained in \(\noending{H}\), the function
\(
G\circ\gamma
\)
is continuous and strictly increasing.

\item For every causal curve \(\gamma:[0,1]\to H\) satisfying
\(
\gamma((0,1))\subset \noending{H}
\),
the quantities
\[
\sup_{t\in(0,1)}G(\gamma_t)
\qquad\text{and}\qquad
\inf_{t\in(0,1)}G(\gamma_t)
\]
are finite.
\end{enumerate}

A gauge function \(G\) is said to be \emph{proper} if, for every compact set \(K\subset\mathbb R\), the preimage
\(
G^{-1}(K)
\)
is precompact in \(\noending{H}\).
\end{definition}

 Assumption (1) implies that, for every causal curve \(\gamma\) contained in \(\noending{H}\), the composition
\(
G\circ\gamma
\)
is continuous and non-decreasing. Indeed, if \(\gamma\) is not
injective, one can first reparametrize it as an injective causal curve
and then apply Assumption~(1).

Assumption (2) prevents  the gauge from blowing up along finite segments of causal curves and ensures that the affine parameter encoded by \(G\) remains finite whenever the underlying causal curve has finite parameter length.

The notion of properness plays a particularly important role in the synthetic theory. As we shall see in Section~\ref{Sec:PenroseC0}, properness of the gauge provides a synthetic counterpart of future null geodesic completeness and constitutes a key ingredient in the formulation of the synthetic Penrose's singularity theorem.
\smallskip

A gauge function allows one to recover, in a purely synthetic manner, the affine structure carried by the null generators of a smooth null hypersurface. This leads to the following synthetic analogue of an affinely parametrized null geodesic.

\begin{definition}[\(G\)-causal curve]
Let \(H\) be a closed achronal set and let
\(
G:\noending{H}\to\mathbb R
\)
be a gauge. A continuous curve
\[
\gamma:[0,1]\to H
\]
is called \emph{\(G\)-causal} if the following conditions hold:

\begin{enumerate}
\item \(\gamma\) is causal;

\item \(\gamma((0,1))\subset\noending{H}\);

\item the function
\(
G\circ\gamma|_{(0,1)}
\)
is affine.
\end{enumerate}

The collection of all \(G\)-causal curves is denoted by
\(
\Caus_G\subset C([0,1];H).
\)
\end{definition}

The third condition should be viewed as the synthetic counterpart of the geodesic equation in the smooth setting. Recall that, for smooth null hypersurfaces, the geodesic equation singles out affine parametrizations of the null generators. In the absence of a differentiable structure, this role is played by the gauge function \(G\): requiring \(G\circ\gamma\) to be affine provides a synthetic notion of affine parameter.

More precisely, if \(H\) is a smooth null hypersurface and \(G\) is induced by an affine parameter along its null generators, then the \(G\)-causal curves are precisely the affinely parametrized null geodesics contained in \(H\). Thus the notion of \(G\)-causal curve extends the classical concept of affinely parametrized null geodesic to the synthetic setting.

\subsubsection{Synthetic null hypersurfaces}

\begin{definition}[Synthetic null hypersurface]\label{D:synthetic-null-hypersurface}
Let \((X,\le,\ll,\mathfrak T)\) be a topological causal space. A triple
\((H,G,\mathfrak m)\) is called a \emph{synthetic null hypersurface} if:

\begin{itemize}
\item \(H\subset X\) is a closed achronal set;

\item \(G:\noending{H}
\to\mathbb R\) is a gauge function;

\item \(\mathfrak m\in\mathcal M^+(H)\) is a non-negative Radon measure that does not charge the static points of \(H\), i.e.\ \(\mm(\staticWithH)=0\).
\end{itemize}
\end{definition}

 Achronality is a fundamental feature of null hypersurfaces.  
 Yet it is far from sufficient to grasp the distinctive features of null geometry,
 in particular curvature.
 Indeed, even spacelike hypersurfaces are achronal, yet they lack the causal generators that constitute the hallmark of null geometry.
 Any synthetic notion of null hypersurface must therefore retain not only achronality, but also a non-trivial causal structure.

The second and third conditions are designed precisely to capture this additional geometric information. The gauge function \(G\) is defined on the non-endpoint part \(\noending{H}\) of \(H\) and encodes the affine structure of its causal generators. At the same time, the requirement
\(
\mathfrak m(\staticWithH)=0
\)
ensures that the reference measure is concentrated on the dynamically relevant part of the hypersurface, namely the points through which non-trivial causal generators pass.

Taken together, the data \((H,G,\mathfrak m)\) provide a synthetic analogue of a smooth null hypersurface endowed with an affine parameter along its null generators and a rigged measure. In particular, they furnish exactly the structures needed to develop a theory of entropy, optimal transport, and synthetic curvature bounds in null directions.

\subsubsection*{Examples arising from smooth spacetimes}

We now discuss several important classes of examples showing that the notion of synthetic null hypersurface naturally encompasses the classical objects of smooth Lorentzian geometry.

\smallskip 
\noindent 
\emph{Smooth null hypersurfaces with a global cross-section}. 
Let \(H\subset M\) be a smooth achronal null hypersurface whose null generators do not form closed causal cycles. Equivalently, the restriction of the causal relation to \(H\) is antisymmetric, so that it defines a partial order on \(H\). Assume moreover that \(H\) admits a smooth global cross-section \(S\subset H\). The existence of \(S\) allows one to construct a global null geodesic vector field \(L\) tangent to \(H\), whose integral curves coincide with the null generators of the hypersurface. Denoting by
\[
\Psi_L:S\times\mathbb R\longrightarrow H
\]
the flow generated by \(L\), every point \(z\in H\) can be uniquely represented in the form
\[
z=\Psi_L(p,t)
\]
for some \(p\in S\) and \(t\in\mathbb R\). This yields a natural gauge function
\[
G_{L,S}:H\longrightarrow\mathbb R,
\]
defined by setting \(G_{L,S}(z)=t\). In other words, \(G_{L,S}\) records the affine parameter of the null generator passing through \(z\), measured relative to the reference cross-section \(S\).

As reference measure on \(H\), one takes the rigged measure \(\mathrm{Vol}_L\) associated with the null geodesic vector field \(L\). The resulting triple
\(
(H,G_{L,S},\mathrm{Vol}_L)
\)
is a synthetic null hypersurface. For more details see~\cite[Rem.\ 3.10]{CMM:25}.

\smallskip 
\noindent 
\emph{Local constructions.} The previous construction can be localized and therefore does not require the existence of a global cross-section. This is particularly important in situations where a global section is unavailable, such as compact Cauchy horizons~\cite{MoncIsenCMP1983, BustReiGRG21, GuMinCMP2022}. In such cases one constructs local gauges and local reference measures, which can then be used to endow suitable open subsets of the hypersurface with the structure of synthetic null hypersurfaces. For more details see~\cite[Rem.\ 3.11]{CMM:25}.

\smallskip
\noindent
\emph{Achronal boundaries.} A particularly significant source of examples is given by achronal boundaries. Let \((M,g)\) be a causally simple \(C^2\) Lorentzian manifold (equivalently, a strongly causal and causally closed spacetime),  and let
\(
A\subset M
\)
be a compact, achronal, spacelike \(C^2\)-submanifold of codimension \(k\ge 2\). Consider the achronal boundary
\[
H=\partial I^+(A).
\]

Although achronal boundaries are generally far from smooth, they retain a remarkable amount of null geometry. Classical results in causality theory show that \(H\) is ruled by null generators and enjoys many of the structural properties of smooth null hypersurfaces; see, for instance,~\cite[Lem.~3.17]{Penrose-DiffTopGR} and~\cite[Prop.~6.3.1]{HawEll}. This residual geometric structure is sufficient to construct, in a natural way, both a gauge function and a reference measure on \(H\).  Thus \(H\) naturally carries the structure of a synthetic null hypersurface (see~\cite[Rem.~3.12]{CMM:25} for more details).

This example is particularly important for applications. Indeed, achronal boundaries naturally appear throughout Lorentzian geometry and general relativity, including in the study of light cones, event horizons, Cauchy horizons, trapped surfaces, and singularity theorems. Consequently, the synthetic framework applies not only to smooth null hypersurfaces but also to many geometrically significant hypersurfaces that arise naturally in low-regularity settings.

\subsection{The synthetic null energy condition \(\mathsf{NC}^{e}(N)\)}

Having introduced synthetic null hypersurfaces, we can now formulate the synthetic counterpart of the null energy condition. As in the smooth theory recalled in Section~\ref{SS:NEC-OT-smooth}, the guiding principle is that curvature should be encoded through displacement-concavity properties of entropy along suitable families of null geodesics. The key difference is that all notions are now expressed purely in terms of the causal structure, the gauge function, and the reference measure, without any appeal to a differentiable structure or curvature tensor.

Let \((H,G,\mathfrak m)\) be a synthetic null hypersurface. Given two probability measures \(\mu_0,\mu_1\in\Prob(H)\), we denote by
\[
\OptGeo^G(\mu_0,\mu_1)
:=
\left\{
\nu\in\Prob(\Caus_G):
(\ee_i)_\sharp\nu=\mu_i,
\quad i=0,1
\right\},
\]
the set of dynamical transport plans concentrated on \(G\)-causal curves with marginals \(\mu_0\) and \(\mu_1\). For any
\(
\nu\in\OptGeo^G(\mu_0,\mu_1),
\)
we write
\[
\mu_t:=(\ee_t)_\sharp\nu,
\qquad
t\in[0,1],
\]
for the associated interpolation. Recall that, given a reference measure \(\mathfrak m\), the entropy-power functional is defined by
\[
\mathcal U_N(\mu|\mathfrak m)
:=
\exp\!\left(
-\frac{1}{N}\Ent(\mu|\mathfrak m)
\right).
\]

The synthetic \(\mathsf{NC}^{e}(N)\) condition introduced in~\cite{CMM:25} is obtained by requiring displacement concavity of \(\mathcal U_{N-1}\) along suitable \(G\)-causal transport plans.

\begin{definition}[Synthetic null energy condition {\cite{CMM:25}}]
Let \(N>0\), let \((X,\ll,\le,\mathfrak T)\) be a topological causal space, and let \((H,G,\mathfrak m)\) be a synthetic null hypersurface.

We say that \((H,G,\mathfrak m)\) satisfies the \emph{synthetic null energy condition} \(\mathsf{NC}^{e}(N)\) if, for every pair of probability measures \(\mu_0,\mu_1\in\Prob(H)\) satisfying
\(
\Pi_{\le}(\mu_0,\mu_1)\neq\emptyset,
\)
there exists a transport plan
\(
\nu\in\OptGeo^G(\mu_0,\mu_1)
\)
such that
\[
\mathcal U_{N-1}(\mu_t|\mathfrak m)
\ge
(1-t)\,
\mathcal U_{N-1}(\mu_0|\mathfrak m)
+
t\,
\mathcal U_{N-1}(\mu_1|\mathfrak m),
\qquad
\forall\, t\in[0,1],
\]
where
\(
\mu_t=(\ee_t)_\sharp\nu.
\)
\end{definition}

This definition should be viewed as the direct synthetic analogue of the smooth characterization discussed in the previous section. In the smooth setting, the null energy condition is equivalent to the displacement concavity of the entropy power along null-geodesic transport plans contained in null hypersurfaces. The above definition takes this displacement-concavity property as the primary object and promotes it to a synthetic curvature condition.

A notable feature of the \(\mathsf{NC}^{e}(N)\) condition is that it only involves the causal structure of the ambient space, the gauge function \(G\), the reference measure \(\mathfrak m\), and entropy functionals. In particular, it makes no reference to a metric tensor, a connection, or Ricci curvature. Consequently, it remains meaningful in settings where none of these differential-geometric objects are available.

As we shall see in the following sections, the \(\mathsf{NC}^{e}(N)\) condition enjoys many of the properties expected of a synthetic curvature notion. In particular, it is compatible with the smooth theory, invariant under natural changes of the reference measure, stable under suitable notions of convergence, and strong enough to imply synthetic counterparts of classical results such as Hawking's area theorem and Penrose's singularity theorem.

\subsection{Fundamental properties of the \(\mathsf{NC}^{e}(N)\) condition}

A synthetic curvature condition is useful only insofar as it enjoys the fundamental properties expected of a geometric notion. In the case of the \(\mathsf{NC}^{e}(N)\) condition, three such properties play a central role:

\begin{itemize}
\item compatibility with the classical smooth null energy condition;

\item well-posedness, namely independence of the auxiliary choices entering the definition of a synthetic null hypersurface;

\item stability under suitable notions of convergence.
\end{itemize}

We briefly discuss each of these aspects.

\subsubsection*{Compatibility with the smooth theory}

The first fundamental requirement is that the synthetic theory recovers the classical one whenever a smooth spacetime structure is available. This is indeed the case, see~\cite[Thms.\ 3.18, 3.19]{CMM24a}: for smooth null hypersurfaces $H$ endowed with their natural gauge function and rigged measure, the synthetic \(\mathsf{NC}^{e}(N)\)  is equivalent to the non-negativity of Ricci tensor in null directions along $H$. Thus the synthetic theory genuinely extends, rather than replaces, the classical smooth framework.

\subsubsection*{Well-posedness}

The definition of a synthetic null hypersurface involves two pieces of auxiliary data: a gauge function \(G\), encoding the affine parametrization of the causal generators, and a reference measure \(\mathfrak m\), playing the role of a rigged measure. A priori, it is therefore not clear whether the resulting \(\mathsf{NC}^{e}(N)\) condition depends on these choices.

To formulate the appropriate notion of invariance, one first identifies the transformations that preserve the underlying null geometry.

Let \(H\) be a closed achronal set. A function
\(
f:\noending{H}\to\mathbb R
\)
is called \emph{transverse} if it is constant along the causal generators of \(H\), that is, if \(f\circ\gamma\) is constant for every causal curve \(\gamma\) contained in \(\noending{H}\).

Using transverse functions, one introduces an equivalence relation on reference measures. Given
\(
\mathfrak m_1,\mathfrak m_2\in\mathcal M^+(H),
\)
we write
\(
\mathfrak m_1\sim\mathfrak m_2
\)
if
\[
\mathfrak m_1\llcorner_{H\setminus\noending{H}}
=
\mathfrak m_2\llcorner_{H\setminus\noending{H}}
\]
and
\[
\mathfrak m_1\llcorner_{\noending{H}}
=
f\,\mathfrak m_2\llcorner_{\noending{H}}
\]
for some positive transverse function \(f\).

Similarly, given two gauges
\(
G_1,G_2:\noending{H}\to\mathbb R,
\)
we write
\(
G_1\sim G_2
\)
whenever they determine the same family of distinguished causal curves, namely
\[
\Caus_{G_1}
=
\Caus_{G_2}.
\]

These equivalence relations have a clear interpretation in the smooth setting. If \(H\) is a smooth null hypersurface, changing the null geodesic vector field \(L\) modifies both the associated rigged measure and the affine parametrization of the null generators. The relations above are precisely the synthetic counterparts of these classical gauge freedoms.

The following result shows that the \(\mathsf{NC}^{e}(N)\) condition is intrinsic.
For the notion of a \emph{tight} gauge see ~\cite[Def.~3.16]{CMM:25}; roughly, it means that the set of $G$-causal curves satisfy a pre-compactness condition \'a-la Arzerl\'a-Ascoli.

\begin{theorem}[{\cite[Thm.\ 4.6]{CMM:25}}]
Let
\(
(H,G_1,\mathfrak m_1)\),
\(
(H,G_2,\mathfrak m_2)
\)
be synthetic null hypersurfaces with tight 
gauges
satisfying
\(
G_1\sim G_2\) and
\(
\mathfrak m_1\sim\mathfrak m_2.
\)
\\Then
\(
(H,G_1,\mathfrak m_1)\) satisfies 
\(
\mathsf{NC}^{e}(N)
\)
if and only if
\(
(H,G_2,\mathfrak m_2)\)
 satisfies 
\(
\mathsf{NC}^{e}(N).
\)
\end{theorem}

Consequently, the synthetic null energy condition depends only on the underlying null geometry and not on the particular gauge function or reference measure used to describe it. In this sense, the condition is geometrically well posed.

\subsubsection*{Stability under convergence}

The third fundamental property is stability. One of the principal motivations for introducing synthetic curvature conditions is their applicability to singular limits of smooth geometries. Accordingly, one expects the \(\mathsf{NC}^{e}(N)\) condition to be preserved under suitable notions of convergence.

This is indeed the case. In~\cite[Sect.~9]{CMM:25}, the authors introduce Lorentzian analogues of pointed measured Gromov{--}Hausdorff convergence and Sturm's \(\mathbb D\)-convergence adapted to the setting of synthetic null hypersurfaces. Roughly speaking, the role traditionally played by the distance function in metric geometry is replaced by the causal and affine structures encoded by the gauge function \(G\). The actual construction is considerably more subtle, however, as it is designed to be invariant under the natural equivalence relation \(G_1\sim G_2\) discussed above.

One of the main results of~\cite{CMM:25} states that the \(\mathsf{NC}^{e}(N)\) condition is stable under these notions of convergence. In particular, limits of synthetic null hypersurfaces satisfying \(\mathsf{NC}^{e}(N)\) continue to satisfy the same condition.

This stability property is one of the major strengths of the optimal-transport approach. It allows one to pass to geometric limits and to formulate meaningful lower bounds on null Ricci curvature for singular spacetime models arising as limits of smooth ones. More generally, it provides a robust framework for studying low-regularity and non-smooth Lorentzian geometries beyond the reach of classical differential-geometric methods.

We conclude by mentioning that, motivated by similar questions, several notions of convergence for smooth and non-smooth spacetimes have recently been proposed in the literature; see, for instance,~\cite{CaMo:20,MiSu:2024,Muller,ByMiSu:2025,SaSo:2025,MoSae:2025}. Understanding the relationship between these convergence theories and the one introduced in~\cite[Sect.~9]{CMM:25} remains an interesting and largely open problem.

\subsection{Penrose singularity theorem in  $C^0$-spacetimes}\label{Sec:PenroseC0}

One of the principal motivations for introducing synthetic versions of the null energy condition is the possibility of extending classical singularity theorems to spacetime settings of very low regularity. Among such results, Penrose's singularity theorem occupies a distinguished position.

Recall that Penrose's original theorem~\cite{Penrose65} asserts that a smooth spacetime satisfying the null energy condition, admitting a non-compact Cauchy hypersurface, and containing a future-trapped surface must be null geodesically incomplete. More precisely, there exists a future-directed null geodesic satisfying
\begin{equation}\label{eq:GeodEq}
\nabla_{\dot\gamma}\dot\gamma=0
\end{equation}
whose maximal domain of definition is not the whole real line.

Over the last decade, substantial progress has been made in extending Penrose's theorem to lower regularity settings. The theorem was established for \(C^{1,1}\)-Lorentzian metrics by Kunzinger, Steinbauer and Vickers~\cite{KSV-CQG}, and subsequently for \(C^1\)-metrics by Graf~\cite{Graf}. In both cases, the null energy condition is interpreted in a distributional sense and geodesics remain classical solutions of an ordinary differential equation.

The case of continuous Lorentzian metrics presents substantially greater difficulties. Indeed, when \(g\in C^0\), neither the Ricci tensor nor the null energy condition can be interpreted distributionally in any natural way. Moreover, the geodesic equation itself is no longer available, and even the notion of a future-trapped surface becomes problematic, since its classical definition involves mean curvature and therefore requires at least \(C^1\)-regularity of the metric.

The synthetic framework recalled in the previous sections provides a way around these difficulties. The basic idea is to replace the classical null energy condition by the synthetic condition \(\mathsf{NC}^{e}(N)\), and to replace null geodesics by \(G\)-causal curves. To formulate a singularity theorem, however, one must still identify appropriate synthetic counterparts of null geodesic completeness and of the trapped-surface condition.

\subsubsection*{Proper gauges and synthetic completeness}

A key observation is that, in the smooth setting, null geodesic completeness imposes a strong constraint on the associated gauge function. More precisely, if every null generator can be extended indefinitely towards the future, then the corresponding gauge is proper.

This suggests taking properness of the gauge as the synthetic analogue of null geodesic completeness. Consequently, a singularity theorem should assert that, under suitable geometric assumptions, the gauge cannot be proper.
This characterization of null completeness is discussed in details in~\cite[Sect.~8.1]{CMM:25}.

\subsubsection*{Synthetic future convergence}

The second issue concerns the trapped-surface assumption.

Classically, the relevant hypothesis is that a compact spacelike surface be future converging, namely that its mean curvature vector points sufficiently towards the future. Since mean curvature depends on the Christoffel symbols, this notion is unavailable for continuous metrics.

The key idea in~\cite[Sect.~7~and~8]{CMM:25} is to replace the classical mean-curvature condition by a second-order variational property involving the reference measure. The motivation comes from the smooth theory: mean curvature arises as the first variation of area, while area itself may be viewed as the first variation of volume. It is therefore natural to seek a synthetic analogue of future convergence in terms of a second variation of volume.

In the synthetic setting, the role of volume is played by the reference measure \(\mathfrak m\). The corresponding notion of area is given by the Minkowski content \(\mathfrak m_G^{+}\), defined with respect to the gauge function \(G\); see~\cite[Def.~7.1]{CMM:25}. This leads to a purely measure-theoretic formulation of future convergence, which remains meaningful even in the absence of differentiable structures and curvature quantities.

Let \((H,G,\mathfrak m)\) be a synthetic null hypersurface satisfying \(\mathsf{NC}^{e}(N)\), and suppose that
\[
H=\partial I^+(S).
\]

\begin{definition}
The set \(S\) is said to be \emph{\((G,\mathfrak m)\)-future converging} if there exists a constant \(\theta<0\) such that
\[
\limsup_{\varepsilon\to 0^+}
\frac{
\mathfrak m(A_\varepsilon^+)
-
\varepsilon\,\mathfrak m_G^+(A)
}{
\varepsilon^2/2
}
\le
\theta\,\mathfrak m_G^+(A)
\]
for every measurable subset \(A\subset S\).
\end{definition}

Although somewhat technical at first sight, this condition admits a natural geometric interpretation: it expresses a strictly negative second-order variation of volume along the future null generators of \(H\), and therefore serves as a synthetic analogue of negative null expansion.

\subsubsection*{Penrose's theorem for continuous Lorentzian metrics}

With these notions in place, one obtains the following synthetic version of Penrose's singularity theorem.
The assumption of \emph{null non-branching}~\cite[Def.~5.6]{CMM:25} is the synthetic counterpart of the uniqueness of null generators in the smooth setting. Roughly speaking, it requires that null generators do not branch, so that two generators that agree on a non-trivial segment must coincide. This condition is automatically verified for achronal sets in \(C^2\)-Lorentzian manifolds; see~\cite[Cor.~5.9]{CMM:25}.

\begin{theorem}[{\cite[Cor.\ 8.8]{CMM:25}}]
Let \((M,g)\) be a spacetime endowed with a continuous Lorentzian metric. Let \(S\subset M\) be a compact achronal set, and let
\[
H=\partial I^+(S).
\]
Assume that \(H\) is endowed with a gauge function \(G\) and a positive Radon measure
\(
\mathfrak m\in\mathcal M^+(H)
\)
with
\(
\operatorname{supp}(\mathfrak m)=H.
\)

Suppose moreover that:

\begin{itemize}
\item \((H,G,\mathfrak m)\) satisfies the \(\mathsf{NC}^{e}(N)\) condition for some \(N>2\);

\item \((M,g)\) admits a non-compact Cauchy hypersurface;

\item \(S\) is \((G,\mathfrak m)\)-future converging;

\item \(H\) is null non-branching.
\end{itemize}

Then the gauge \(G\) cannot be proper.
\end{theorem}

The conclusion should be interpreted as a synthetic form of null geodesic incompleteness. Indeed, one of the key observations underlying the theorem is that, in the smooth setting, future null geodesic completeness of \(H\) implies properness of the associated gauge function. Therefore, the failure of properness can be viewed as the synthetic counterpart of the existence of an incomplete null generator.

Consequently, when specialized to smooth spacetimes, the theorem recovers the qualitative conclusion of Penrose's classical singularity theorem. More importantly, it remains meaningful for merely continuous Lorentzian metrics, a regularity regime in which neither Ricci curvature, nor the geodesic equation, nor mean curvature are available in their classical forms.

The theorem therefore provides a striking illustration of the strength of the optimal transport approach: by replacing differential-geometric notions with synthetic entropy-concavity conditions, one can recover one of the deepest results of general relativity in a setting far beyond the reach of classical methods.

\bibliographystyle{acm}
\bibliography{literature.bib}

\begin{thebibliography}{10}

\bibitem{AGKS:23}
{\sc Alexander, S.~B., Graf, M., Kunzinger, M., and S{\"a}mann, C.}
\newblock Generalized cones as {Lorentzian} length spaces: causality,
  curvature, and singularity theorems.
\newblock {\em Commun. Anal. Geom. 31}, 6 (2023), 1469--1528.

\bibitem{Amb:ICM}
{\sc Ambrosio, L.}
\newblock Calculus, heat flow and curvature-dimension bounds in metric measure
  spaces.
\newblock In {\em Proceedings of the international congress of mathematicians
  2018, ICM 2018, Rio de Janeiro, Brazil, August 1--9, 2018. Volume I. Plenary
  lectures}. Hackensack, NJ: World Scientific; Rio de Janeiro: Sociedade
  Brasileira de Matem{\'a}tica (SBM), 2018, pp.~301--340.

\bibitem{AES:16}
{\sc Ambrosio, L., Erbar, M., and Savar{\'e}, G.}
\newblock Optimal transport, {Cheeger} energies and contractivity of dynamic
  transport distances in extended spaces.
\newblock {\em Nonlinear Anal., Theory Methods Appl., Ser. A, Theory Methods
  137\/} (2016), 77--134.

\bibitem{AGMR}
{\sc Ambrosio, L., Gigli, N., Mondino, A., and Rajala, T.}
\newblock {R}iemannian {R}icci curvature lower bounds in metric measure spaces
  with {$\sigma$}-finite measure.
\newblock {\em Trans. Amer. Math. Soc. 367}, 7 (2015), 4661--4701.

\bibitem{AGS14a}
{\sc Ambrosio, L., Gigli, N., and Savar{\'e}, G.}
\newblock Calculus and heat flow in metric measure spaces and applications to
  spaces with {R}icci bounds from below.
\newblock {\em Inventiones mathematicae 195}, 2 (2014), 289--391.

\bibitem{AGS14b}
{\sc Ambrosio, L., Gigli, N., and Savar{\'e}, G.}
\newblock Metric measure spaces with {R}iemannian {R}icci curvature bounded
  from below.
\newblock {\em Duke Mathematical Journal 163}, 7 (2014), 1405--1490.

\bibitem{AGS15}
{\sc Ambrosio, L., Gigli, N., and Savar{\'e}, G.}
\newblock {B}akry--{\'{e}}mery curvature-dimension condition and {R}iemannian
  {R}icci curvature bounds.
\newblock {\em The Annals of Probability 43}, 1 (2015), 339--404.

\bibitem{AMS19}
{\sc Ambrosio, L., Mondino, A., and Savar{\'e}, G.}
\newblock Nonlinear diffusion equations and curvature conditions in metric
  measure spaces.
\newblock {\em Memoirs of the American Mathematical Society 262}, 1270 (2019).

\bibitem{Octet}
{\sc Beran, T., Braun, M., Calisti, M., Gigli, N., McCann, R.~J., Ohanyan, A.,
  Rott, F., and Sämann, C.}
\newblock A nonlinear d'{A}lembert comparison theorem and causal differential
  calculus on metric measure spacetimes.
\newblock Preprint at arXiv:{2408.15968v2}, 2025.

\bibitem{BHS:26}
{\sc Beran, T., Harvey, J., and S{\"a}mann, C.}
\newblock Curvature bounds, regularity and inextendibility of spacetimes.
\newblock Preprint, {arXiv}:2603.20802 [gr-qc] (2026), 2026.

\bibitem{Braun:exact}
{\sc Braun, M.}
\newblock Exact d'{Alembertian} for {Lorentz} distance functions.
\newblock Preprint, {arXiv}:2408.16525, to appear in {\it Calc. Var. PDE}.

\bibitem{Braun}
{\sc Braun, M.}
\newblock Rényi's entropy on {L}orentzian spaces. {T}imelike
  curvature-dimension conditions.
\newblock {\em J. Math. Pures Appl. (9) 177\/} (2023), 46--128.

\bibitem{BSC:Lipsch}
{\sc Braun, M., and Candal, M.~S.}
\newblock Comparison theory for {Lipschitz} spacetimes.
\newblock Preprint, {arXiv}:2603.24195 [math.{DG}] (2026), 2026.

\bibitem{Braun-McCann}
{\sc Braun, M., and McCann, R.~J.}
\newblock Causal convergence conditions through variable timelike {Ricci}
  curvature bounds.
\newblock Preprint, {arXiv}:2312.17158, to appear in {\it Mem.\ Europ.\ Math.\
  Soc.}

\bibitem{BR:GL}
{\sc Braun, M., and Rotolo, C.}
\newblock A synthetic {Gannon}-{Lee} incompleteness theorem.
\newblock Preprint, {arXiv}:2602.14246 [gr-qc] (2026), 2026.

\bibitem{BuLe}
{\sc Burtscher, A.~Y., and LeFloch, P.~G.}
\newblock The formation of trapped surfaces in spherically-symmetric
  {E}instein--{E}uler spacetimes with bounded variation.
\newblock {\em J. Math. Pures Appl. (9) 102}, 6 (2014), 1164--1217.

\bibitem{Busemann}
{\sc Busemann, H.}
\newblock Timelike spaces.
\newblock {\em Diss. Math. 53\/} (1967).

\bibitem{BustReiGRG21}
{\sc Bustamante, I., and Reiris, M.}
\newblock A classification theorem for compact {Cauchy} horizons in vacuum
  spacetimes.
\newblock {\em Gen. Relativ. Gravitation 53}, 3 (2021), 10.
\newblock Id/No 36.

\bibitem{ByMiSu:2025}
{\sc Bykov, A., Minguzzi, E., and Suhr, S.}
\newblock Lorentzian metric spaces and {GH}-convergence: the unbounded case.
\newblock {\em Lett. Math. Phys. 115}, 3 (2025), Paper No. 63, 66.

\bibitem{CGHKS-25}
{\sc Calisti, M., Graf, M., Hafemann, E., Kunzinger, M., and Steinbauer, R.}
\newblock Hawking's singularity theorem for {Lipschitz} {Lorentzian} metrics.
\newblock {\em Commun. Math. Phys. 406}, 9 (2025), 31.
\newblock Id/No 207.

\bibitem{Carroll}
{\sc Carroll, S.~M.}
\newblock {\em Spacetime and geometry. {An} introduction to general
  relativity}.
\newblock Cambridge University Press, 2019.

\bibitem{CMM24a}
{\sc Cavalletti, F., Manini, D., and Mondino, A.}
\newblock Optimal transport on null hypersurfaces and the null energy
  condition.
\newblock {\em Commun. Math. Phys. 406}, 9 (2025), 62.
\newblock Id/No 212.

\bibitem{CMM:25}
{\sc Cavalletti, F., Manini, D., and Mondino, A.}
\newblock On the geometry of synthetic null hypersurfaces.
\newblock Preprint, {arXiv}:2506.04934 [math.{DG}] (2026), 2026.

\bibitem{CaMi:21}
{\sc Cavalletti, F., and Milman, E.}
\newblock The globalization theorem for the curvature-dimension condition.
\newblock {\em Invent. Math. 226}, 1 (2021), 1--137.

\bibitem{CM:22}
{\sc Cavalletti, F., and Mondino, A.}
\newblock A review of {L}orentzian synthetic theory of timelike {R}icci
  curvature bounds.
\newblock {\em Gen. Relativity Gravitation 54}, 11 (2022), Paper No. 137, 39.

\bibitem{CaMo:20}
{\sc Cavalletti, F., and Mondino, A.}
\newblock Optimal transport in {L}orentzian synthetic spaces, synthetic
  timelike {R}icci curvature lower bounds and applications.
\newblock {\em Camb. J. Math. 12}, 2 (2024), 417–534.

\bibitem{CM24-IsopLor}
{\sc Cavalletti, F., and Mondino, A.}
\newblock A sharp isoperimetric-type inequality for {L}orentzian spaces
  satisfying timelike {R}icci lower bounds, 2024.
\newblock Preprint at arXiv:2401.03949v2.

\bibitem{CBCMG-2009}
{\sc Choquet-Bruhat, Y., Chru{\'s}ciel, P.~T., and Mart{\'{\i}}n-Garc{\'{\i}}a,
  J.~M.}
\newblock The light-cone theorem.
\newblock {\em Classical Quantum Gravity 26}, 13 (2009), 135011 (22pp).

\bibitem{Chr-Grant}
{\sc Chru{\'s}ciel, P.~T., and Grant, J. D.~E.}
\newblock On {Lorentzian} causality with continuous metrics.
\newblock {\em Classical Quantum Gravity 29}, 14 (2012), 32.
\newblock Id/No 145001.

\bibitem{CEMS}
{\sc Cordero-Erausquin, D., McCann, R.~J., and Schmuckenschl\"ager, M.}
\newblock A {R}iemannian interpolation inequality \`a{} la {B}orell, {B}rascamp
  and {L}ieb.
\newblock {\em Invent. Math. 146}, 2 (2001), 219--257.

\bibitem{Curiel2017}
{\sc Curiel, E.}
\newblock {\em A Primer on Energy Conditions}.
\newblock Springer New York, New York, NY, 2017, pp.~43--104.

\bibitem{DCT-1991}
{\sc Dembo, A., Cover, T.~M., and Thomas, J.~A.}
\newblock Information-theoretic inequalities.
\newblock {\em IEEE Trans. Inform. Theory 37}, 6 (1991), 1501--1518.

\bibitem{EM17}
{\sc Eckstein, M., and Miller, T.}
\newblock Causality for nonlocal phenomena.
\newblock {\em Ann. Henri Poincar\'{e} 18}, 9 (2017), 3049--3096.

\bibitem{EKS}
{\sc Erbar, M., Kuwada, K., and Sturm, K.-T.}
\newblock On the equivalence of the entropic curvature-dimension condition and
  {B}ochner's inequality on metric measure spaces.
\newblock {\em Invent. Math. 201}, 3 (2015), 993--1071.

\bibitem{FewsterGalloway2011}
{\sc Fewster, C.~J., and Galloway, G.~J.}
\newblock Singularity theorems from weakened energy conditions.
\newblock {\em Classical and Quantum Gravity 28}, 12 (2011), 125009.

\bibitem{FewsterSmith2008}
{\sc Fewster, C.~J., and Smith, C.~J.}
\newblock Absolute quantum energy inequalities in curved spacetime.
\newblock {\em Annales Henri Poincar\'e 9\/} (2008), 425--455.

\bibitem{Ford1978}
{\sc Ford, L.~H.}
\newblock Quantum coherence effects and the second law of thermodynamics.
\newblock {\em Proceedings of the Royal Society A 364}, 1717 (1978), 227--236.

\bibitem{GeTr}
{\sc Geroch, R., and Traschen, J.}
\newblock Strings and other distributional sources in general relativity.
\newblock {\em Phys. Rev. D (3) 36}, 4 (1987), 1017--1031.

\bibitem{Gigli15}
{\sc Gigli, N.}
\newblock On the differential structure of metric measure spaces and
  applications.
\newblock {\em Memoirs of the American Mathematical Society 236}, 1113 (2015),
  vi+91.

\bibitem{G:Survey}
{\sc Gigli, N.}
\newblock De {Giorgi} and {Gromov} working together.
\newblock Preprint, {arXiv}:2306.14604 [math.{MG}] (2023), 2023.

\bibitem{GMS15}
{\sc Gigli, N., Mondino, A., and Savar\'{e}, G.}
\newblock Convergence of pointed non-compact metric measure spaces and
  stability of {R}icci curvature bounds and heat flows.
\newblock {\em Proc. Lond. Math. Soc. (3) 111}, 5 (2015), 1071--1129.

\bibitem{Graf}
{\sc Graf, M.}
\newblock Singularity theorems for {$C^1$}-{L}orentzian metrics.
\newblock {\em Comm. Math. Phys. 378}, 2 (2020), 1417--1450.

\bibitem{Grant}
{\sc Grant, J. D.~E.}
\newblock Areas and volumes for null cones.
\newblock {\em Ann. Henri Poincar\'{e} 12}, 5 (2011), 965--985.

\bibitem{GKS:19}
{\sc Grant, J. D.~E., Kunzinger, M., and S{\"a}mann, C.}
\newblock Inextendibility of spacetimes and {Lorentzian} length spaces.
\newblock {\em Ann. Global Anal. Geom. 55}, 1 (2019), 133--147.

\bibitem{GuMinCMP2022}
{\sc Gurriaran, S., and Minguzzi, E.}
\newblock Surface gravity of compact non-degenerate horizons under the dominant
  energy condition.
\newblock {\em Commun. Math. Phys. 395}, 2 (2022), 679--713.

\bibitem{Rigging}
{\sc Guti\'{e}rrez, M., and Olea, B.}
\newblock Induced {R}iemannian structures on null hypersurfaces.
\newblock {\em Math. Nachr. 289}, 10 (2016), 1219--1236.

\bibitem{Haw:67}
{\sc Hawking, S.~W.}
\newblock The occurrence of singularities in cosmology. {III}: {Causality} and
  singularities.
\newblock {\em Proc. R. Soc. Lond., Ser. A 300\/} (1967), 187--201.

\bibitem{Haw71}
{\sc Hawking, S.~W.}
\newblock Black holes in general relativity.
\newblock {\em Commun. Math. Phys. 25\/} (1972), 152--166.

\bibitem{HawEll}
{\sc Hawking, S.~W., and Ellis, G. F.~R.}
\newblock {\em The large scale structure of space-time}.
\newblock Cambridge Monographs on Mathematical Physics. Cambridge University
  Press, 1973.

\bibitem{HawkingPenrose1970}
{\sc Hawking, S.~W., and Penrose, R.}
\newblock The singularities of gravitational collapse and cosmology.
\newblock {\em Proceedings of the Royal Society A 314\/} (1970), 529--548.

\bibitem{Ket24}
{\sc Ketterer, C.}
\newblock Characterization of the null energy condition via displacement
  convexity of entropy.
\newblock {\em Journ. London Math. Soc. 109}, 1 (2024), e12846 (24pp).

\bibitem{KontouFewster2020}
{\sc Kontou, E.-A., and Fewster, C.~J.}
\newblock Generalized quantum energy conditions in curved spacetime.
\newblock {\em Classical and Quantum Gravity 37}, 19 (2020), 193001.

\bibitem{CausalSpace}
{\sc Kronheimer, E.~H., and Penrose, R.}
\newblock On the structure of causal spaces.
\newblock {\em Proc. Cambridge Philos. Soc. 63\/} (1967), 481--501.

\bibitem{KS}
{\sc Kunzinger, M., and S\"{a}mann, C.}
\newblock Lorentzian length spaces.
\newblock {\em Ann. Global Anal. Geom. 54}, 3 (2018), 399--447.

\bibitem{KSV-CQG}
{\sc Kunzinger, M., Steinbauer, R., and Vickers, J.~A.}
\newblock The {Penrose} singularity theorem in regularity {{\(C^{1,1}\)}}.
\newblock {\em Classical Quantum Gravity 32}, 15 (2015), 12.
\newblock Id/No 155010.

\bibitem{LLV}
{\sc Le~Floch, B., LeFloch, P.~G., and Veneziano, G.}
\newblock Cyclic spacetimes through singularity scattering maps. {T}he laws of
  quiescent bounces.
\newblock {\em J. High Energy Phys.}, 4 (2022), Paper No. 95, 72.

\bibitem{LeFlochMardare2007}
{\sc LeFloch, P.~G., and Mardare, C.}
\newblock Definition and weak stability of lorentzian manifolds with
  distributional curvature.
\newblock {\em Portugaliae Mathematica 64\/} (2007), 535--573.

\bibitem{Lichn}
{\sc Lichnerowicz, A.}
\newblock {\em Th\'eories relativistes de la gravitation et de
  l'\'electromagn\'etisme. {R}elativit\'e{} g\'en\'erale et th\'eories
  unitaires}.
\newblock Masson et Cie, Paris, 1955.

\bibitem{lottvillani}
{\sc Lott, J., and Villani, C.}
\newblock Ricci curvature for metric-measure spaces via optimal transport.
\newblock {\em Ann. of Math. (2) 169}, 3 (2009), 903--991.

\bibitem{MaSe}
{\sc Mars, M., and Senovilla, J. M.~M.}
\newblock Geometry of general hypersurfaces in spacetime: junction conditions.
\newblock {\em Classical Quantum Gravity 10}, 9 (1993), 1865--1897.

\bibitem{McCann}
{\sc McCann, R.~J.}
\newblock Displacement convexity of {B}oltzmann's entropy characterizes the
  strong energy condition from general relativity.
\newblock {\em Camb. J. Math. 8}, 3 (2020), 609--681.

\bibitem{McCann-NEC}
{\sc McCann, R.~J.}
\newblock A synthetic null energy condition.
\newblock {\em Commun. Math. Phys. 405}, 2 (2024), 38 (24pp).

\bibitem{MinguzziCMP15}
{\sc Minguzzi, E.}
\newblock Area theorem and smoothness of compact {C}auchy horizons.
\newblock {\em Commun. Math. Phys. 339}, 1 (2015), 57--98.

\bibitem{Min}
{\sc Minguzzi, E.}
\newblock Causality theory for closed cone structures with applications.
\newblock {\em Rev. Math. Phys. 31}, 5 (2019), 1930001, 139.

\bibitem{Minguzzi-Review2019}
{\sc Minguzzi, E.}
\newblock Lorentzian causality theory.
\newblock {\em Living Rev. Relativ. 22\/} (2019), 202.
\newblock Id/No 3.

\bibitem{Minguzzi-Suhr}
{\sc Minguzzi, E., and Suhr, S.}
\newblock Lorentzian metric spaces and their {Gromov}-{Hausdorff} convergence.
\newblock {\em Lett. Math. Phys. 114}, 3 (2024), 73 (63pp).

\bibitem{MiSu:2024}
{\sc Minguzzi, E., and Suhr, S.}
\newblock Lorentzian metric spaces and their {Gromov}-{Hausdorff} convergence.
\newblock {\em Lett. Math. Phys. 114}, 3 (2024), 63.
\newblock Id/No 73.

\bibitem{Gravitation}
{\sc Misner, C.~W., Thorne, K.~S., and Wheeler, J.~A.}
\newblock {\em Gravitation}.
\newblock Princeton University Press, 2017.

\bibitem{MoncIsenCMP1983}
{\sc Moncrief, V., and Isenberg, J.}
\newblock Symmetries of cosmological {Cauchy} horizons.
\newblock {\em Commun. Math. Phys. 89\/} (1983), 387--413.

\bibitem{MRS:IPPW}
{\sc Mondino, A., Ryborz, V., and S{\"a}mann, C.}
\newblock Stability of {Synthetic} {Timelike} {Ricci} {Bounds} under
  {$C^0$}-{Limits} and {Applications} to {Impulsive} {Gravitational} {Waves}.
\newblock Preprint, {arXiv}:2605.03172 [gr-qc] (2026), 2026.

\bibitem{MoSu}
{\sc Mondino, A., and Suhr, S.}
\newblock An optimal transport formulation of the {E}instein equations of
  general relativity.
\newblock {\em J. Eur. Math. Soc. 25}, 3 (2022), 933--994.

\bibitem{MoSae:2025}
{\sc Mondino, A., and Sämann, C.}
\newblock Lorentzian gromov–hausdorff convergence and pre-compactness.
\newblock {\em Journal für die reine und angewandte Mathematik (Crelles
  Journal). {\rm doi:10.1515/crelle-2026-0067}\/} (2026).

\bibitem{Muller}
{\sc Müller, O.}
\newblock Gromov-{H}ausdorff metrics and dimensions of {L}orentzian length
  spaces.
\newblock Preprint at arXiv:2209.12736, 2024.

\bibitem{Nachbin}
{\sc Nachbin, L.}
\newblock Topology and order.
\newblock Van {Nostrand} {Mathematical} {Studies}. {Vol}. 4.
  {Princeton}-{New}-{Jersey}-{Toronto}-{New} {York}-{London}: {D}. {Van}
  {Nostrand} {Company}, {Inc}. 122 p. (1965)., 1965.

\bibitem{OnSC:Orlicz}
{\sc Ohanyan, A., and Candal, M.~S.}
\newblock Timelike {Ricci} curvature lower bounds via optimal transport for
  {Orlicz}-type {Lorentzian} costs.
\newblock Preprint, {arXiv}:2604.22538 [math.{DG}] (2026), 2026.

\bibitem{O'Neill}
{\sc O'Neill, B.}
\newblock {\em Semi-{R}iemannian geometry, with applications to relativity},
  vol.~103 of {\em Pure and Applied Mathematics}.
\newblock Academic Press, New York, 1983.

\bibitem{OttoVillani}
{\sc Otto, F., and Villani, C.}
\newblock Generalization of an inequality by {T}alagrand and links with the
  logarithmic {S}obolev inequality.
\newblock {\em J. Funct. Anal. 173}, 2 (2000), 361--400.

\bibitem{Penrose65}
{\sc Penrose, R.}
\newblock Gravitational collapse and space-time singularities.
\newblock {\em Phys. Rev. Lett. 14\/} (1965), 57--59.

\bibitem{PenGW}
{\sc Penrose, R.}
\newblock The geometry of impulsive gravitational waves.
\newblock General {Relativity}, {Papers} {Honour} {J}. {L}. {Synge}, 101-115
  (1972)., 1972.

\bibitem{Penrose-DiffTopGR}
{\sc Penrose, R.}
\newblock {\em Techniques of differential topology in relativity}, vol.~7 of
  {\em CBMS-NSF Reg. Conf. Ser. Appl. Math.}
\newblock Society for Industrial {and} Applied Mathematics (SIAM),
  Philadelphia, PA, 1972.

\bibitem{Penrose:GRG02F}
{\sc Penrose, R.}
\newblock Gravitational collapse: {The} role of general relativity.
\newblock {\em Gen. Relativ. Gravitation 34}, 7 (2002), 1141--1165.

\bibitem{SaSo:2025}
{\sc Sakovich, A., and Sormani, C.}
\newblock Introducing various notions of distances between space-times.
\newblock Preprint at arXiv:2410.16800, 2025.

\bibitem{SaC0}
{\sc S\"{a}mann, C.}
\newblock Global hyperbolicity for spacetimes with continuous metrics.
\newblock {\em Ann. Henri Poincar\'{e} 17}, 6 (2016), 1429--1455.

\bibitem{Sen:98}
{\sc Senovilla, J. M.~M.}
\newblock Singularity theorems and their consequences.
\newblock {\em Gen. Relativ. Gravitation 30}, 5 (1998), 701--848.

\bibitem{sturm:I}
{\sc Sturm, K.-T.}
\newblock On the geometry of metric measure spaces. {I}.
\newblock {\em Acta Math. 196}, 1 (2006), 65--131.

\bibitem{sturm:II}
{\sc Sturm, K.-T.}
\newblock On the geometry of metric measure spaces. {II}.
\newblock {\em Acta Math. 196}, 1 (2006), 133--177.

\bibitem{Sturm:ECM}
{\sc Sturm, K.-T.}
\newblock Metric measure spaces and synthetic {Ricci} bounds: fundamental
  concepts and recent developments.
\newblock In {\em European congress of mathematics. Proceedings of the 8th
  congress, 8ECM, Portoro\v{z}, Slovenia, June 20--26, 2021}. Berlin: European
  Mathematical Society (EMS), 2023, pp.~125--159.

\bibitem{Suhr}
{\sc Suhr, S.}
\newblock Theory of optimal transport for {L}orentzian cost functions.
\newblock {\em M\"{u}nster J. Math. 11}, 1 (2018), 13--47.

\bibitem{Vil:BS}
{\sc Villani, C.}
\newblock Isoperimetric inequalities in metric measure spaces [after {F}.
  {Cavalletti} \& {A}. {Mondino}].
\newblock In {\em S\'eminaire Bourbaki. Volume 2016/2017. Expos\'es
  1120--1135}. Soci{\'e}t{\'e} Math{\'e}matique de France (SMF), 2019, p.~ex.

\bibitem{vRS}
{\sc von Renesse, M.-K., and Sturm, K.-T.}
\newblock Transport inequalities, gradient estimates, entropy, and {R}icci
  curvature.
\newblock {\em Comm. Pure Appl. Math. 58}, 7 (2005), 923--940.

\bibitem{Wald}
{\sc Wald, R.~M.}
\newblock {\em General relativity}.
\newblock University of Chicago Press, 1984.

\end{thebibliography}
\end{document}